\newcommand{\CLASSINPUTtoptextmargin}{0.76in}
\newcommand{\CLASSINPUTbottomtextmargin}{1in}
\documentclass[conference]{IEEEtran}
\IEEEoverridecommandlockouts
\usepackage{amsmath,amssymb,amsfonts}
\usepackage{graphicx}
\usepackage{textcomp}
\usepackage{xcolor}
\def\BibTeX{{\rm B\kern-.05em{\sc i\kern-.025em b}\kern-.08em
    T\kern-.1667em\lower.7ex\hbox{E}\kern-.125emX}}

\usepackage[english]{babel}
\usepackage[utf8]{inputenc}
\usepackage[noend]{algpseudocode}
\usepackage{lipsum}
\usepackage{fullwidth}

\usepackage{microtype}
\usepackage[skins]{tcolorbox}
\usepackage{wrapfig}
\usepackage{threeparttable}
\usepackage{diagbox}
\usepackage{multirow}
\usepackage{subfigure}
\usepackage{float}
\usepackage{graphicx} 
\usepackage{xspace}
\usepackage{enumitem}
\usepackage{booktabs}
\usepackage{xcolor}
\usepackage{hyperref}
\usepackage{ragged2e}
\usepackage{pifont}

\usepackage{nicematrix}
\usepackage{bibspacing}

\newcommand{\zhengqing}[1]{{\bf\color{red} [zhengqing: #1]}}

\usepackage[linesnumbered,ruled,vlined]{algorithm2e}
\usepackage{algpseudocode}  

\newtheorem{theorem}{Theorem}[section]
\newtheorem{proposition}[theorem]{Proposition}

\newcommand{\nop}[1]{}
\newcommand{\ours}{HMES\xspace}
\newtheorem{problem}{Problem}

\newcommand{\tabincell}[2]{\begin{tabular}{@{}#1@{}}#2\end{tabular}}

\usepackage{marvosym}

\begin{document}

\title{HMES: A Scalable Human Mobility and Epidemic Simulation System with Fast Intervention Modeling
}
\makeatletter
\newcommand{\linebreakand}{%
  \end{@IEEEauthorhalign}
  \hfill\mbox{}\par
  \mbox{}\hfill\begin{@IEEEauthorhalign}
}
\makeatother

\author{
\IEEEauthorblockN{Haoyu Geng, Guanjie Zheng\textsuperscript{\Letter}\thanks{Guanjie Zheng is the corresponding author}, Zhengqing Han}
\IEEEauthorblockA{
\textit{Shanghai Jiao Tong University}\\
genghaoyu98,gjzheng,14-hz@sjtu.edu.cn}
\and
\IEEEauthorblockN{Hua Wei}
\IEEEauthorblockA{
\textit{New Jersey Institute of Technology}\\
hua.wei@njit.edu}
\and 
\IEEEauthorblockN{Zhenhui Li}
\IEEEauthorblockA{
\textit{The Pennsylvania State University}\\
jessieli@ist.psu.edu}
}

\IEEEaftertitletext{\vspace{-2.5\baselineskip}}

\maketitle

\begin{abstract}
Recently, the world has witnessed the most severe pandemic (COVID-19) in this century. Studies on epidemic prediction and simulation have received increasing attention. However, the current methods suffer from three issues. First, most of the current studies focus on epidemic prediction, which can not provide adequate support for intervention policy making. Second, most of the current interventions are based on population groups rather than fine-grained individuals, which can not make the measures towards the infected people and may cause waste of medical resources. Third, current simulations are not efficient and flexible enough for large-scale complex systems.

In this paper, we propose a new epidemic simulation framework called \ours to address above three challenges. The proposed framework covers a full pipeline of epidemic simulation and enables comprehensive fine-grained control in large scale. In addition, we conduct experiments on real COVID-19 data. \ours demonstrates more accurate modeling of disease transmission up to 300 million people and up to 3 times acceleration compared to the state-of-the-art methods. 
\end{abstract}


\section{Introduction}\label{sec:introduction}
\vspace{-0.15cm}
The year 2020 has witnessed one of the most severe worldwide pandemic, coronavirus (COVID-19) in this century, which has caused more than 600 million confirmed cases\footnote{Data accessed from \url{https://coronavirus.jhu.edu/map.html} on 08/28/2022}. Recently, many researches~\cite{kdd2018-intracity,kdd19-epideep,song2020reinforced,bengio2020predicting,hao2020understanding} have been conducted on various aspects of pandemic, including peak date and height prediction, estimation of R0(reproduction number), influence of factors like mobility, etc. These studies aim to discover more about the pandemic (i.e., how it propagates, how to prepare) before the pandemic happens.


In this paper, we would focus on \textbf{simulation} for epidemic control, aiming at providing an environment for intervention policy development and evaluation. This line of simulation studies is significantly distinct from the line of works on epidemic prediction \cite{bisset2009modeling,kdd2018-intracity,kdd19-epideep,song2020reinforced,bengio2020predicting,hao2020understanding}, where only the final infected count or peak time is given. This is because that current epidemic prediction studies cannot handle the evaluation of intervention policies (e.g., vaccination, isolation), and hence can not help to mitigate pandemic directly. In addition, the epidemic prediction relies heavily on the data distribution (determined by the collection location and time), the generalization of these methods remains doubtful. In contrast, introducing simulation methods can bring further explainability and accuracy support.

\looseness=-1 Towards simulation for epidemic control, we are faced with following challenges.
First, building \textit{fine-grained intervention} for a large-scale complex urban system is difficult. Note that, a city may owns a population up to 20$\sim$30 million, while in previous works in epidemic intervention\cite{kdd2018-intracity,hao2020understanding,lorch2020quantifying,bengio2020predicting,chang2021mobility,chang2021supporting}, cities are usually segmented into hundreds of groups. Since simulating the epidemic progress and intervention on individual level instead of groups involves much higher time and space complexity, it remains unsolved how to simulate the control pandemic among 20 million individuals. Still, individual-based simulation has the following compelling strengths making worth the large computing cost. (1) Compared with group-level epidemic control, individual-level control exhibits better effectiveness in preventing disease propagation. (2) Precise individual-level control would preserve the freedom of people to the most extent and greatly reduce cost on administrative and medical resources.

\looseness=-1 Second, generating the individual-level epidemic control policy requires extensive computation over the enormous individual contact tracing process. For a city with population density $q$ (can be as large as 20 million people over thousands of tracts) and $|\mathbf{F}|$ infected people, tracing the contact of the infected people can be as complex as $O(|\mathbf{F}|q)$, which is impractical for real use, not to mention if second or higher-order contact tracing are needed. Therefore, it is highly desirable that more scalable contact tracing methods are proposed and finally make the individual-level precise control come true.

\nop{
\textbf{Scalable} and \textbf{flexible} simulation design is required to make our come true, where current methods on epidemic simulation~\cite{song2020reinforced,lorch2020quantifying} would not be available to this challenge as they could only fit small-scale data and would not generalize to arbitrary settings.}

Therefore, in this paper, we propose a new modular epidemic simulation system called \ours (\textbf{H}uman \textbf{M}obility and \textbf{E}pidemic \textbf{S}imulation), which support fine-grained intervention control.  First, to overcome the complex modeling process in epidemic simulation, we propose a \textbf{flexible}  modular pipeline that could incorporate variants for mobility, epidemic modules and a customized intervention interface. Modules can be replaced at need. In this paper, we demonstrate the example implementation for modules in framework. Second, we improve the \textbf{scalabiltity} to large scale by reducing memory overhead and running time by two key insights.
To begin with, we adopt regularity-based human mobility~\cite{gonzalez2008understanding,brockmann2006scaling,mcinerney2013breaking,pappalardo2018data,stanley2018many} to reduce the redundancy in human mobility simulation.
Moreover, we design a improved contact tracing so that the redundant tracing is saved. We further provide equivalence guarantee and algorithm design that conducts minimum necessary tracing without extra time complexity. 
\begin{figure}[htbp]
    \centering
    \begin{tabular}{c}
    \includegraphics[width=0.46\textwidth]{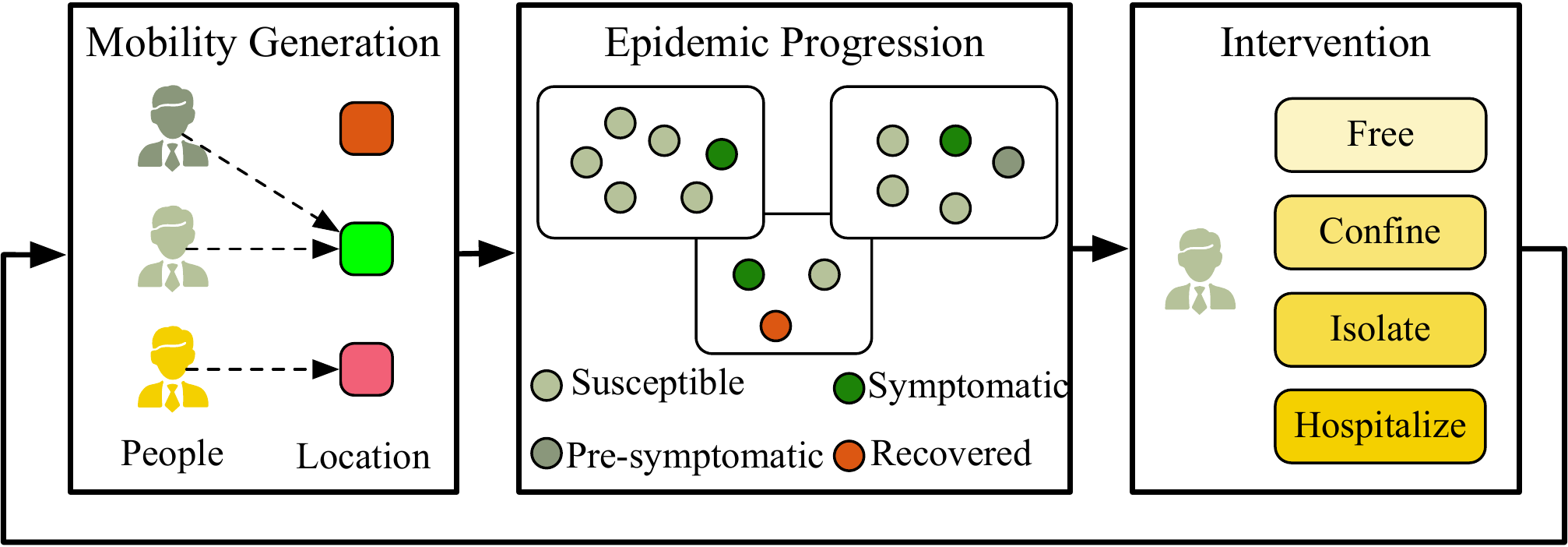}
    \end{tabular}
    \caption{Complete framework of \ours, consist of three consecutive modules}
    \label{fig:ours}
    \vspace{-0.4cm}
\end{figure}

\looseness=-1 Compared with previous, \ours is different in three aspects. 
\begin{itemize}[leftmargin=*]
    \item We build a simulation \textit{framework} that supports \textbf{flexible} modular design. \ours can not only provide interface where the modules could be replaced by substitutions, but also support intervention strategies with user-customized operations (both individual-level and group-level). Among these, we propose minimum necessary location tracing without extra computational complexity.
    \item We demonstrate \textbf{accurate} simulation of the disease transmission, which achieves better performance than previous epidemic prediction and simulation baselines.
    \item We deploy an \textbf{efficient} framework by efficient contact tracing with up to 0.3 billion population and 3 times faster in runtime, by consideration on mobility regularity and reduction on bipartite contact tracing tree.
\end{itemize}
\Urlmuskip=0mu  plus 10mu
We provide Python package with easy-following use cases as well as our original C implementation\footnote{\url{https://github.com/hygeng/humanflow}}.
We hope this work would boost further research in epidemic modeling with more complex, effective, and humanistic intervention policies to mitigate the pandemic. 


\section{Related Work}
\subsection{Simulation for Epidemic Propagation}
\vspace{-3pt}
Current epidemic simulations could be categorized as two types:  \textbf{microscopic} and \textbf{macroscopic}, where the former build the mobility for each individual and simulate the epidemic propagation when people contact with each other, and the latter simulate pandemic propagation in sub-population groups.

{\bf Microscopic Approaches}
could be divided into location-based and social-network based. 
Location-based methods are widely-adopted methods to build people-location bipartite graphs in urban simulations.
Episims~\cite{episims} first proposes dynamic bipartite graphs to model the physical contact patterns that result from movements of individuals between specific locations
Further, EpiSimdemics~\cite{episimdemics}  proposes a interaction-based simulation. 
Currently, FRED~\cite{grefenstette2013fred} is the most recent model that supports epidemic simulation for every state in the US taking health behavior patterns and mixing pattern of population into account. 
EpidemicSim~\cite{EpidemicSim} builds the mobility pattern as individuals walking on a defined space, and implements the epidemic simulation by statistical models. but only support simulation of small scale.
\cite{gupta2020covi} recently proposed COVIAgentSim, an agent-based compartmental simulator. \cite{bengio2020predicting} further adopts it to propose proactive contact tracing to predict an individual’s infectiousness (risk of infecting others) based on their contact history and other information. However, these methods are based on small-size groups, which is incapable of supporting city-wise simulation and applications upon it.

\looseness=-1 For another approach, social-network-based simulation traces historical contacted people in social networks.
EpiFast~\cite{epifast} is classical method that builds the social contact network for epidemic simulation. However, without the bipartite graph between people and locations, Epifast 
fails to support additional functions like tracing contacts. 
FastSIR~\cite{fastsir} is the most recent method which proposes an efficient recursive method for calculating the probability distribution of the number of infected nodes. Average case running time are reduced compared to the naive SIR model in FastSIR method.
STAND~\cite{xu2019stand} builds the diffusion of contagions in networks in a general view with a probalistic spatial-temporal process. 

{\bf Macroscopic Approaches}
often analyze statistical features for epidemic modeling.
The epidemic spread on arbitrary complex networks is well studied in SIR (susceptible, Infected, Recovered) model \cite{1927SIR}. Later, several variations of the SIR models are proposed, including SEIR~\cite{biswas2014seir}, SIRS~\cite{2010sirs}. Most simulators build their epidemic propagation with these methods and their variants mentioned above. 
\cite{chang2021mobility} is a recent work on mobility and epidemic simulation on the mapping of subpopulation groups to POI on a large scale, and studies the intervention policies of specific POIs. However, without individual-based study on intervention, it only provides overall statistical estimates without fine-grained suggestions to individuals.
There are also some recent papers that build simulations for COVID-19 \cite{chang2021supporting,hao2020understanding}.


However, the macroscopic models lack mobility with individual granularity, which constrain their further improvements. Current social-network-based simulation takes pre-defined social network for each person which saves mobility generation, while the running time of pair-wise epidemic propagation is exponentially slow, making it fail in expanding to large scale.
In this work, we adopt the microscopic method to build a dynamic bipartite graph to support individual-level intervention strategies. We take location-based approach and propose algorithms to speed up the whole process to support urban simulation of mobility, epidemic propagation, and interventions on large scale.

\vspace{-6pt}
\subsection{Prediction for Epidemic Propagation and Intervention}
\vspace{-5pt}
There also exists a rich body of literature on epidemic modeling and investigation of various intervention strategies since the outbreak of COVID-19 pandemic.  A paucity of research works has been conducted to analyze the properties and characteristics of  epidemic transmission for COVID-19 in medical and biostatistical views \cite{browne2021infection,flaxman2020report,li2020early}. Some previous works have investigated the influence of intervention strategies towards propagation of COVID-19 through econometric methods and micro-simulation~\cite{hsiang2020effect,nussbaumer2020quarantine,ferguson2020report,bengio2020predicting}.  Some researches start by analyzing real-world mobility data and connect these observations towards the propagation of COVID-19 \cite{kraemer2020effect,chinazzi2020effect,luo2020deeptrack}. Plenty of works have investigated the estimated outcomes with respect to different intervention strategies 
\cite{arenas2020modeling,arenas2020mathematical,flaxman2020estimating}
in a coarse granularity through prediction and estimation.
However, most of the above are empirical analysis and specific case investigations. They may fit well with specific areas and periods, but can not provide support for intervention policy evaluation and development.

\looseness=-1 We also have to note that the goal of prediction tasks is \textbf{orthogonal to us} as they focus on harnessing existing data to fit the epidemic model and analyzing the intervention outcomes through estimation rather than building the real-world simulation. Although there is analysis for various intervention strategies, most of them are based on coarse-grained computation and parameterized estimation rather than individual-based simulation.  In addition, our simulation framework is able to incorporate and support these epidemic and intervention variants as components in our modular design and the specific epidemic and intervention models are not our focus.

\section{Overview}\label{sec:problem-definition}
\vspace{-5pt}

\newlength\q
\setlength\q{\dimexpr .315\textwidth -2\tabcolsep}
\newlength\qq
\setlength\qq{\dimexpr .14\textwidth -2\tabcolsep}
\begin{table}[tp]\label{tab:notations}
\caption{Notations.}
\begin{center}
\begin{threeparttable}
\vspace{-0.3cm}
\noindent\begin{tabular}{p{\q}|p{\qq}}
\toprule
Variable & Notation\\ \midrule
Day index & $d$, ranges from $1$ to $D$ \\
Hour index & $h$, ranges from $1$ to $H$ \\
Simulation step index & $t(t=(d-1) \times H + h)$ \\
Temporal/spatial randomness\tnote{1} & $r$ \\
People set, Location set and index(in bracket)\tnote{2} & $\mathbf{M}(m)$, $\mathbf{L}(l)$  \\
Trajectory of person $m$  at step $t$ (location $l$)& \multirow{1}{*}{$\mathbf{J}_{m}^{t}$ = $l$}\\ 
People set that visited location $l$ at step $t$ 
&  \multirow{1}{*}{$\mathbf{V}_{l}^{t}$}\\
Infection rate between two people, in location $l$ & $p$, $p_l$  \\
Infected, susceptible, recovered and newly infected people set at location $l$ at step $t$  & \multirow{2}{*} {$\mathbf{I}_l^t$, $\mathbf{S}_l^t$, $\mathbf{R}_l^t$, $\mathbf{F}_l^t$}\\
Intervention strategy at day $d$\tnote{3} & $\mathbf{
\lambda^{(d)}}$ \\
Population density & $q$ \\
\bottomrule
\end{tabular}

\begin{tablenotes}
\footnotesize
\item[1] Randomness is probability that individual deviate routine in temporal or spatial view;
\item[2] $|\cdot|$ represents the size of a set. $\sigma(|L|)$ represents the time complexity for generating a random integer within range $[0, |L|)$; 
\item[3] Letters with upper script $t$ means the quantity for time step $t$, e.g., $\mathbf{I}^t$. When necessary, we will change the superscript $t$ to $(d)$ (with bracket) to represent the quantity for day $d$, e.g., $\mathbf{\lambda}^{(d)}$.
\end{tablenotes}
\end{threeparttable}
\end{center}
\vspace{-0.8cm}
\end{table}

\subsection{Problem Definition}
\vspace{-5pt}
The epidemic simulation problem can be defined as follows. Used key notations are summarized in Table~\ref{tab:notations}.
\begin{problem}
Assume a city with location set $\mathbf{L}$, people set $\mathbf{M}$, and period of time to simulate $D$. Human mobility are constrained according to strategy $\lambda$. The simulation aims to do the following three steps.
(1) Generate mobility trajectory $\mathbf{J}_m^t$ for each person $m$ at time step $t$.
(2) Generate newly infected people set $\mathbf{F}^t$ for each time step $t$.
(3) Provide daily intervention strategy $\lambda^{(d)} = f(\mathbf{F}^{(d)}, \mathbf{J}_m^t)$.
\end{problem}

Note that, the temporal granularity of the simulation is a short time interval (e.g., 1 hour), while the intervention strategy are usually made on a daily basis due to the delay of gathering the infection information and conducting intervention in the real world.

\vspace{-3pt}
\subsection{Framework}
\vspace{-5pt}
Our simulation system (as in Figure~\ref{fig:ours}) consists of three modules (Mobility Generation, Epidemic Progression, and Intervention). In general, our simulation pipeline is described in Alg.~\ref{alg:ours}, where the system simulates Mobility Generation and Epidemic Progression in every time step $t$ and conducts Intervention every day $d$. Note that as we are proposing a framework whose modules could be substituted by counterparts(including recent and future modules), we will introduce one example implementation of each module.

\setlength{\textfloatsep}{0pt}
\begin{algorithm}[ht]
  \caption{Pipeline of \ours}  
  \label{alg:ours}  
         \KwIn{People set $\mathbf{M}$, Location set $\mathbf{L}$;  
      Initial infected people set $\mathbf{I}^0$, initial intervention strategy $\lambda^{(0)}$;}  
    \For{$ d = 0 \to D-1$}
    {
        \For{$ h = 0 \to H-1$}
        {
            $t = (d, h)$ \\
             $\mathbf{J}, \mathbf{V} \gets $ \Call{MobilityGeneration}{$\mathbf{M}, \mathbf{L}, \mathbf{\lambda^{(d)}}, t$} \\
             $\mathbf{F^t} \gets $ \Call{EpidemicProgression}{$\mathbf{M},\mathbf{L}, \mathbf{V}, t$}\\
             
            Update $\mathbf{I}, \mathbf{S}, \mathbf{R} $ for each location\\
        }
        Update $\mathbf{F}^{(d)}$ with $\mathbf{F}^t$ \\
        $\lambda^{(d+1)} \gets $ \Call{Intervention}{$\mathbf{F}^{(d)}, \mathbf{J},\mathbf{V}, t$}
    } 
\end{algorithm} 
\vspace{-15pt}
\section{Epidemic Simulation}\label{sec:method}
\vspace{-3pt}
The disease transmission happens when people contact with each other. Hence, in this section, we introduce how to simulate human mobility and afterwards epidemic propagation.
\subsection{Mobility Generation}\label{sec:mobility-generation}
\vspace{-3pt}
The goal of this module is to generate human mobility trajectories. 
Existing epidemic simulations assume people move randomly among pre-selected locations~\cite{episimdemics,EpidemicSim,bisset2009modeling,bhatele2017massively}. 
However, this kind of mobility generation ignores the mobility regularity and will induce unnecessary repetitive computations, and we resolve this by utilizing the mobility regularity to improve the mobility generation algorithm. Before that, we first show some evidence about human mobility regularity.

\textbf{Observation: Human Moves Regularly.}
Studies have shown that people tend to visit a few locations frequently and keep similar schedules every day ~\cite{stanley2018many,brockmann2006scaling,mcinerney2013breaking,pappalardo2018data}.
This kind of regularity can be further illustrated in a case study. Let us categorize locations as three major types: residential, working, and commercial areas. As shown in Figure~\ref{fig:pattern}, a typical individual may frequently visit one residential location, one working location, and several commercial locations nearby. Assume that people's activities are majorly between 7am and 9pm. During weekdays, one individual may leave home for work at time $t_1$, go to mall after work at $t_2$, and return home at $t_3$. As illustrated in Figure~\ref{fig:pattern}, $t_1$, $t_2$ and $t_3$ will usually not vary significantly among different weekdays. 
\begin{figure}[h]
    \centering
    \begin{tabular}{c}
    \includegraphics[width = 0.46\textwidth]{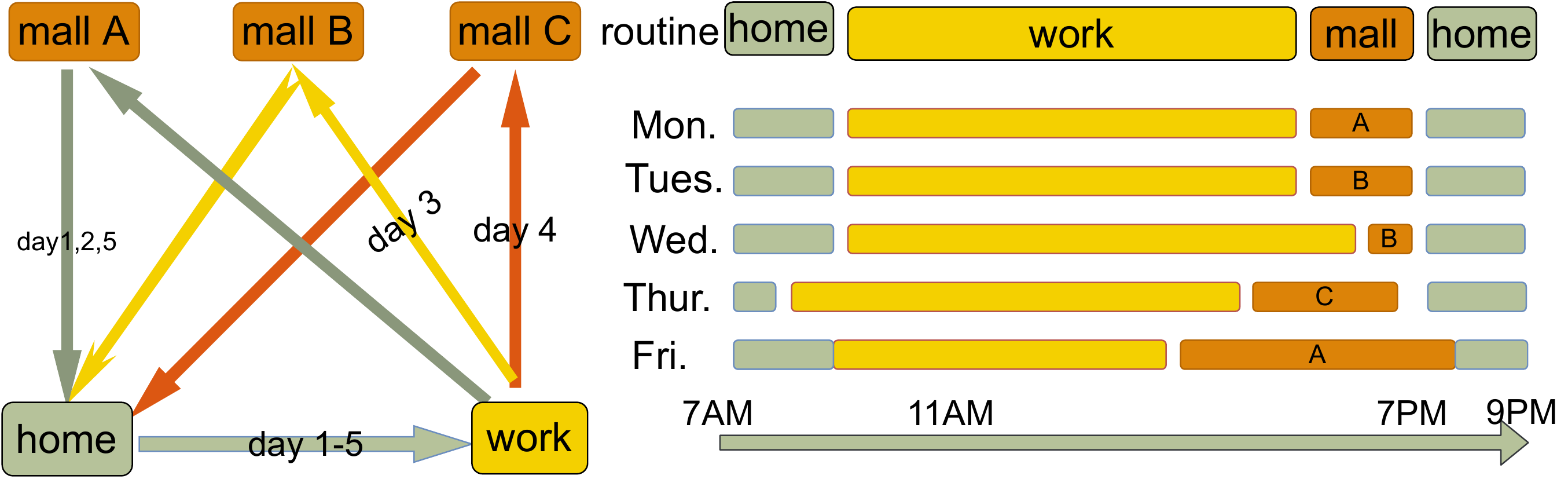}
    \end{tabular}
    \caption{Mobility pattern in spatial \& temporal view}
    \label{fig:pattern}
\end{figure}

\looseness=-1 We further observe how much time people spend in total at their top favorite locations of GeoLife dataset\footnote{Data accessed from www.microsoft.com/en-us/research/project/geolife-building-social-networks-using-human-location-history/} in Figure~\ref{fig:geolife}(a). For instance, the left-most bar means that 50\% of the population has spent more than 90\% time at their top-3 favorite sites. The bar chart shows that a high proportion of people follow strong mobility regularity in their daily pattern. Figure~\ref{fig:geolife}(b) shows the real trajectories of 182 anonymous individuals in multiple days. Darker color means this individual has followed this route more frequently. We observe that people tend to keep visiting their favorite locations. 
\begin{figure}[htbp]
  \vspace{-4mm}
  \centering 
  \subfigure[Percentage of time spent in top-3 favourite locations]{ 
    \includegraphics[width = 0.21\textwidth]{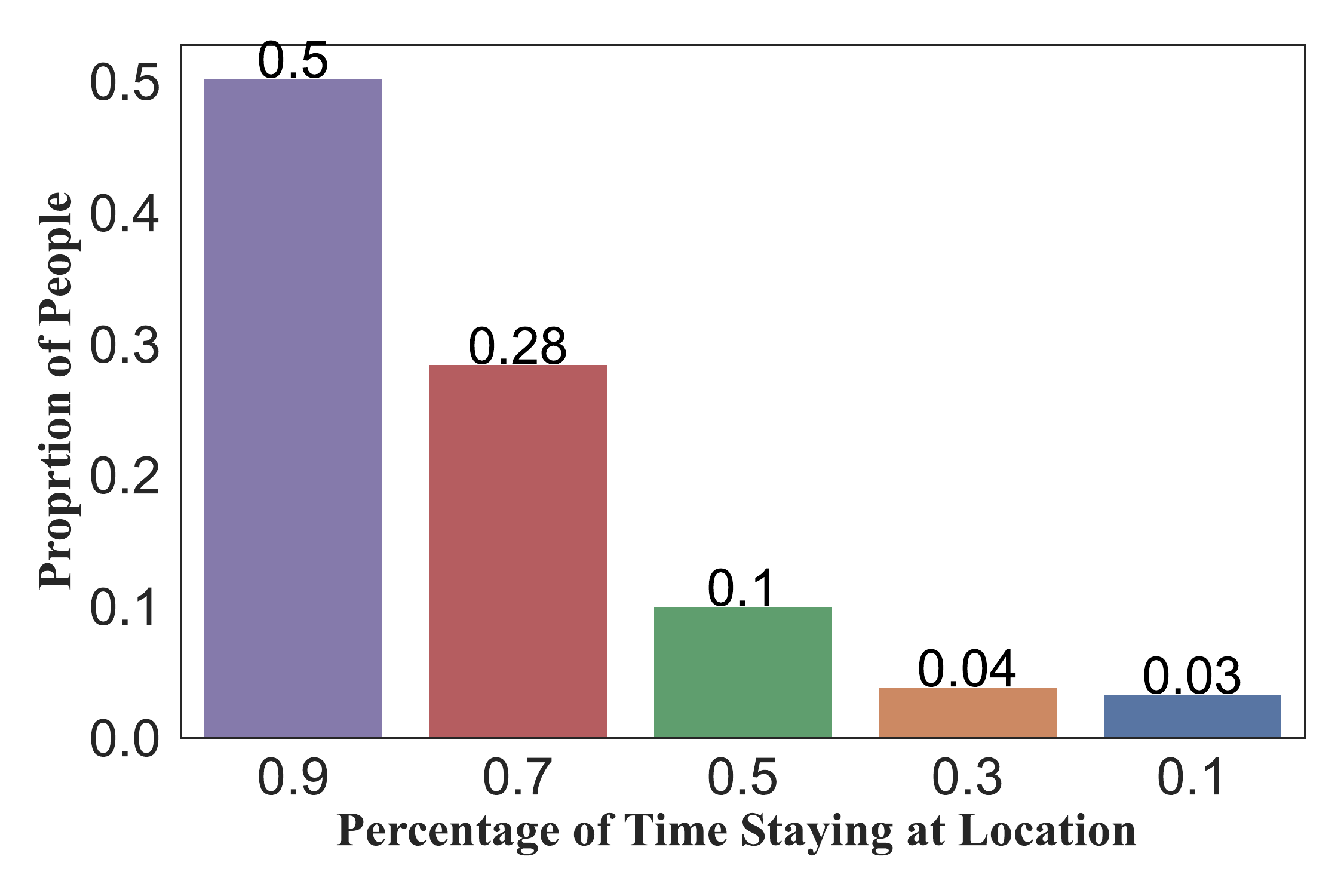}
  } 
  \hfill
  \subfigure[Trajectory visualization of 182 people]{ 
    \includegraphics[width = 0.225\textwidth]{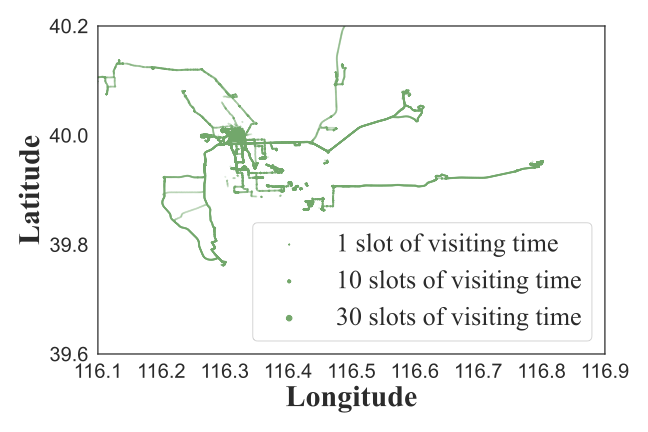}
  } 
  \vspace{-3.8mm}
  \caption{Mobility pattern on GeoLife datase} 
  \label{fig:geolife}
  \vspace{-3mm}
\end{figure}

\textbf{Regularity-based Mobility Generation}
The aforementioned analysis have shown that the modeling of human mobility pattern by regular routines and occasional exceptions are reasonable. Based on this intuition, we propose our mobility generation module to reduce the computation cost. 

We use an incrementally-updated bipartite graph of people and locations to store the mobility trajectory, where each edge indicates a visit of one individual to a location. As demonstrated in Figure~\ref{fig:pattern}, 
different from previous methods which generate and store each edge separately, we construct a template one-day trajectory (consisting of $H$ steps) for each individual and only conduct further generation and storage when this individual deviates from his/her template trajectory. The probability that people deviate from routine trajectory is set as hyper-parameter $r$.

For each time step, if the person follows the routine trajectory, which counts for most circumstances, extra efforts to trace human mobility is saved. Otherwise, this individual will be assigned a random location.  The time complexity of algorithm is $|\mathbf{M}|[r(\sigma(|L|) + O(1)) + (1-r) O (1)]$. The mobility interface is also available for more variants. 
The detailed algorithms and complexity analysis are available in Appendix.

\nop{
As our \ours framework takes incremental trajectory updates, when the users follow their routines, their mobility updates are saved. There is no recording under regularity in the best case, i.e. $A = 0$, where $r_T = r_L = 0 $. When mobility is totally stochastic where $r_T = r_L = 1 $, our model gets its worst case and $A = D \cdot(\sigma (\frac{|\mathbf{L}|}{3})+ H\cdot |\mathbf{M}|)$. Take the analysis from \ref{sec:mobi-pattern}, we take $r_L = 0.2, r_T = 0.1$ as the general case, which covers majority of people.  Time complexity under our proposed speedup $O(|\mathbf{M}| \cdot (3 \cdot \sigma (\frac{|\mathbf{L}|}{3})) + |\mathbf{M}| \cdot D\cdot \left[ 0.2 \cdot \sigma (\frac{ |\mathbf{L}|}{3}) + 0.44H\right])$, while that without speedup (in EpiSimdemics~\cite{episimdemics}) is $A_0$ (without initialization term) listed above. Our speedup could avoid destination assignment for each day as well as the extra records for each hour. }

\vspace{-3pt}
\subsection{Epidemic Progression}
\vspace{-5pt}
The epidemic progression model consists of two parts, within-host progression and between-host progression. Combined with the mobility model, this model will simulate how the disease can spread among different people and locations. Likewise, epidemic component could be replaced with counterparts easily and we give example implementation.

{\bf Within-host Progression}
This part describes how the disease evolves with respect to one individual. Similar to previous studies~\cite{biswas2014seir}, the health status of individuals are represented as four categories: \textit{susceptible, pre-symptomatic, symptomatic, and recovered}. Note that pre-symptomatic and symptomatic people are infectious in different stages and the pre-symptomatic cannot be discovered directly because they do not exhibit observable symptoms. Susceptible  people have a certain chance of being infected when having contact with infectious people. Once an individual is infected, his/her health status transfers from susceptible to pre-symptomatic, and then to symptomatic after an incubation period. People who get infected cannot recover without medical treatment by intervention of hospitalization. Once recovered, they become immune to the disease for long.  Note these settings (e.g., incubation period, immunity) can be easily revised as needed. 

{\bf Between-host Progression}
The disease transmission among different individuals happens when people gather at one location. For each location, first determine the sub-population count of the susceptible, the infected and recovered. Then the probability $p_l$ that one susceptible person get infected at location $l$ can be determined as a function of the above counts. Afterwards, calculate the number of newly infected count $|\mathbf{F}_l^t|$ and randomly select these many people from the susceptible people set. Here, $p$ is the infection rate through two men from the infected to the susceptible. This progress can be described in detail by Alg.3 in Appendix.

\vspace{-3pt}
\section{Intervention Strategies}\label{sec:intervention}
\vspace{-5pt}
\begin{figure*}[tbp] 
\centering 
\includegraphics[width=0.93\textwidth]{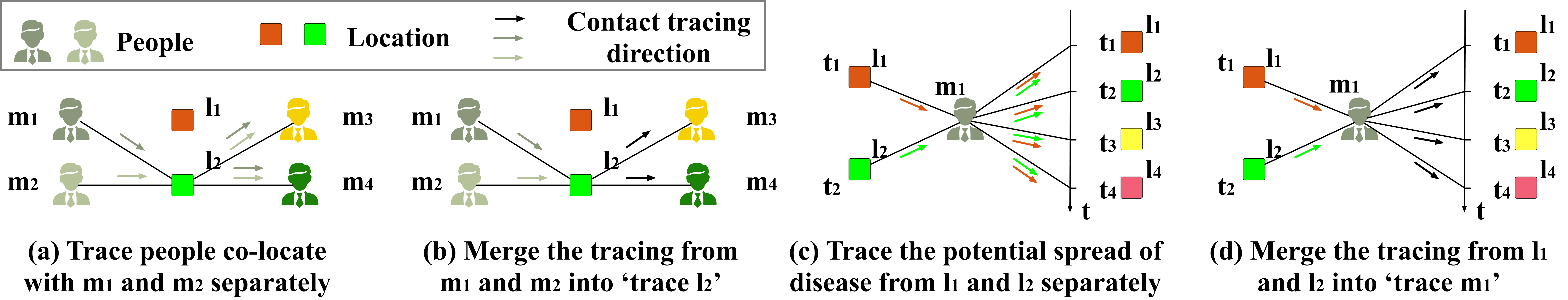} 
\vspace{-0.4cm}
\caption{Contact tracing and pruning. In figure (a) and (c), multiple tracing iterations are done represented by arrows in corresponding color. In figure (b) and (d), unnecessary iterations are saved.} 
\label{fig:fast-intervention}
\vspace{-0.6cm}
\end{figure*}

\setlength{\textfloatsep}{0pt}
\SetInd{0.5em}{0.5em}
\begin{algorithm}[t]  
    \caption{Fast Contact Tracing: Bipartite-Check-Tree}  
    \label{alg:intervention}  
    \KwIn{Daily infected people set $\mathbf{\mathbf{F}^{(d)}}$, people's trajectories  $\mathbf{J}$, visiting people set of locations $\mathbf{V}$, current step $t$}
    \KwOut{Intervention strategy for next day $\mathbf{\lambda^{(d+1)}}$}
    \textbf{Parameters}: Tracking steps $\tau$,  maximum tracing order $maxOrder$, intervention type $\beta$\\
    Initialize potential infection source people set $\mathbf{B}    \gets \mathbf{\mathbf{F^{(d)}}}$ \\
    \For{$ order = 1 \to maxOrder$}{
        \tcc{Update risky locations} 
        Initialize the concerned time-location pair set $\mathbf{C}_L \gets \{\} $\\
            \For{$ m \in \mathbf{B} $}
            {
                \For{$ t' = t - \tau \to t$}{
                    $l = \mathbf{J}_{m}^{t'}$ \\
                    $\mathbf{C}_L \gets \mathbf{C}_L + (t',l)$
                }
            }
        Reset $\mathbf{B} \gets \{\}$ \\
        \tcc{Find minimal  location tracing set $C_{L}^{\star}$} 
        \For{$ t' = t - \tau \to t$}{
            $C_{L,t'}^{\star}$ = Hungarian($C_{L,t'}$)
            }
        \tcc{Update risky people} 
        \For{$ (t',l) \in \mathbf{C}_{L}^{\star} $}{
                \For {$m \in \mathbf{V}_{l}^{t'}$}
                {
                $\mathbf{B} \gets \mathbf{B} + m$;
                    
                }
        }
        
        \For{$ m \in \mathbf{B}$}
        {
             Impose intervention strategy $\mathbf{\lambda}^{(d+1)}[m] \gets \beta$
        }
    }
\end{algorithm}

To mitigate the epidemic progression, people have proposed various strategies, e.g., vaccination and separation. In this paper, we mainly discuss the separation strategy implemented upon mobility and epidemic propagation, the main strategy usually in the early stage of a pandemic outbreak. 

The key of separation strategy is to find the people that may become an infection source. Previously-discussed group-based intervention strategies \cite{kdd2018-intracity,kdd19-epideep,song2020reinforced} aim to find the location/subpopulation with the highest risk and constrain the mobility of this whole group (or part of it). Conversely,  \ours adopt individual-level separation strategies:
Not only individual-level policies are more flexible than group-level policies, it realizes fine-grained epidemic control that reduces the negative impact of the intervention to a minimal scale. We propose a fast contact tracing paradigm to accelerate the process of finding minimum set of risky people.
\nop{
We also propose an individual health risk prediction paradigm to balance effectiveness and cost under limited medical resources with Appendix as extensions.}




{\bf Fast Contact Tracing Query Support}
The basic contact tracing algorithm conducts a two-hop query: (1) find out the places that a concerned person (whose contact needs to be traced) has visited; (2) find out the group of people that have been in the same location with the concerned person. The algorithm is elaborated in the Appendix as Alg.~\ref{alg:intervention-naive}.

\looseness=-1 However, this algorithm has unnecessary repeated computation, as shown in Figure~\ref{fig:fast-intervention} (a)). Apparently, the tracing person $m1$ and $m2$ will lead to the same location $l2$ and then lead to repetitive checking if simply following aforementioned algorithm. Same thing happen when checking one individual for multiple times as in Figure ~\ref{fig:fast-intervention} (c).
The key to accelerate is to ensure checking each individual or location only once. Thus, we propose the Bipartite-Check-Tree algorithm as Alg.~\ref{alg:intervention}. Suppose we want to trace the contacts up to $maxOrder$-th hop of the current infected people. The whole contact tracing process can be regarded as a loop of tracing risky individuals (Figure~\ref{fig:fast-intervention}(b)) and tracing risky locations (Figure~\ref{fig:fast-intervention}(d)). Therefore, the whole tracing process of each infected individual will form a tree with risky individuals in odd layers and risky locations in even layers. Thus, to avoid repetitive checking, we will conduct layer-wise checking and merge the node of risky locations and individuals before entering the next layer. This way, we will achieve minimum times of checking and greatly improve the efficiency of contact tracing.


{\bf Towards minimal location tracing}
The detailed fast contact tracing algorithm is shown in Alg.~\ref{alg:intervention}. We provide an equivalence guarantee on the fast contact tracing algorithm:
\begin{proposition}\label{prop:vertex-cover}
Alg.~\ref{alg:intervention} is capable of grasping the minimal necessary set of locations for contact tracing with reduced time complexity compared with the full contact tracing algorithm.
\end{proposition}
where Hungarian algorithm~\cite{kuhn1955hungarian} is adopted to achieve this goal, which avoid redundant location tracing for those do not contain infected people.
The time complexity of that in Alg.~\ref{alg:intervention} is $O(|\mathbf{B}| \cdot \tau + |\mathbf{C}_L| \cdot q) $. The proof is available in Appendix.
\nop{
Since our fast contact tracing in the intervention aggregates the risky locations, compared to the naive version of complexity shown in $C_0$ above, our algorithm improves from the multiplication of risky location size and  people size to linear addition. Suppose for each person $m \in \mathbf{M}$, the number that he/she has $m$ has co-occurred in the historical trajectories with all other people is $r_m$, then our improvement has saved the query of risky locations that presents co-occurrence. Therefore, the reduced time complexity is $maxOrder \cdot \tau \cdot (\sum_{m\in \mathbf{E}} r_m) \cdot q$, where $maxOrder$ is the order of contact tracing.}

\vspace{-1pt}
\section{Experiments}\label{sec:experiment}
\vspace{-3pt}
We first evaluate the effectiveness of \ours with accurate simulation in Sec.~\ref{sec:experi:effectiveness}, and then demonstrate efficiency in Sec.~\ref{sec:experi:efficiency}, the effects of different intervention strategies in Sec.~\ref{sec:experi:intervene-effectiveness}, and our interface with case study in Sec.~\ref{sec:case study}.

\begin{table*}[t]
\caption{Performance of simulating infection counts of 15 varied-size cities (1-5 indicates city index) in terms of absolute error (number of infected cases) and relative error (in bracket). The lower the better.}
\label{tab:infection-case15}
\vspace{-0.25cm}
\resizebox{\textwidth}{!}{%
\begin{tabular}{l|ccccc|ccccc|ccccc}
\toprule
\multirow{2}{*}{Methods} & \multicolumn{5}{c|}{Small City Index}                  & \multicolumn{5}{c|}{Medium City Index}                 & \multicolumn{5}{c}{Large City Index}                 \\ \cline{2-16} 
                         & 1      & 2       & 3      & 4      & 5      & 1      & 2      & 3      & 4       & 5      & 1      & 2      & 3      & 4      & 5      \\ \midrule
\multirow{2}{*}{HMES}    & \textbf{4}      & \textbf{3}       & \textbf{15}     & 30     & \textbf{31}     & \textbf{104}    & \textbf{513}    & 166    & \textbf{958}     & \textbf{47}     & \textbf{52}     & \textbf{16}     & \textbf{2}      & 247    & \textbf{583}    \\
                         & (2.9\%)  & (3.0\%)   & (13.6)\% & (36.6\%) & (44.9\%) & (5.7\%)  & (22.5\%) & (8.5\%)  & (66.5\%)  & (3.2\%)  & (5.2\%)  & (91.6\%)  & (0.2\%)  & (19.8\% )& (30.1\%) \\ \hline
\multirow{2}{*}{EpiSimdemics}  & 22     & 27      & 33     & 56     & 60     & 1,004  & 535    & 299    & 1,041   & 956    & 80     & 186    & 889    & 298    & 802    \\
                         & (15.8\%) & (26.7\%)  & (30.0\%) & (68.3\%) & (87.0\%) & (54.6\%) & (23.5\%) & (36.4\%) & (72.3\%)  & (65.0\%) & (8.1\%)  & (18.1\%) & (92.7\%) & (23.9\%) & (41.5\%) \\ \hline
\multirow{2}{*}{FRED}    & 28     & 35      & 58     & \textbf{4}      & 41     & 1,660  & 1,133  & 291    & 2,920   & 1,039  & 829    & 121    & 51     & 354    & 1,743  \\
                         & (20.1\%) & (34.7\%)  & (52.7\%) & (4.9\%)  & (59.4\%) & (90.3\%) & (49.8\%) & (14.9\%) & (202.8\%) & (70.6\%) & (83.7\%) & (11.8\%) & (5.3\%)  & (28.3\%) & (90.1\%) \\ \hline
\multirow{2}{*}{Epifast} & 15     & 114     & 26     & 6      & 33     & 1,415  & 648    & \textbf{31}     & 2,438   & 278    & 898    & 62     & 38     & 515    & 1,505  \\
                         & (10.8\%) & (112.9\%) & (23.6\%) & (7.3\%)  & (47.8\%) & (77.0\%) & (28.5\%) & (1.6\%)  & (169.3\%) & (18.9\%) & (90.6\%) & (6.0\%)  & (4.0\%)  & (41.2\%) & (77.8\%) \\ \hline
\multirow{2}{*}{FastSIR} & 9      & 118     & 40     & 105    & 85     & 1,575  & 1,529  & 703    & 2,361   & 2,932  & 895    & 359    & 257    & \textbf{177}    & 583    \\
 & (6.5\%) & (116.8\%) & (36.4\%) & (128.0\%) & (123.2\%) & (85.7\%) & (67.2\%) & (189.3\%) & (164.0\%) & (199.3\%) & (90.3\%) & (34.9\%) & (26.8\%) & (14.2\%) & (30.1\%) \\ \bottomrule
\end{tabular}%
}
\vspace{-2mm}
\end{table*}




\begin{figure*}[tbhp]
    \vspace{-0.2cm}
    \centering
    \begin{tabular}{cccc}
        \includegraphics[width=0.23\textwidth]{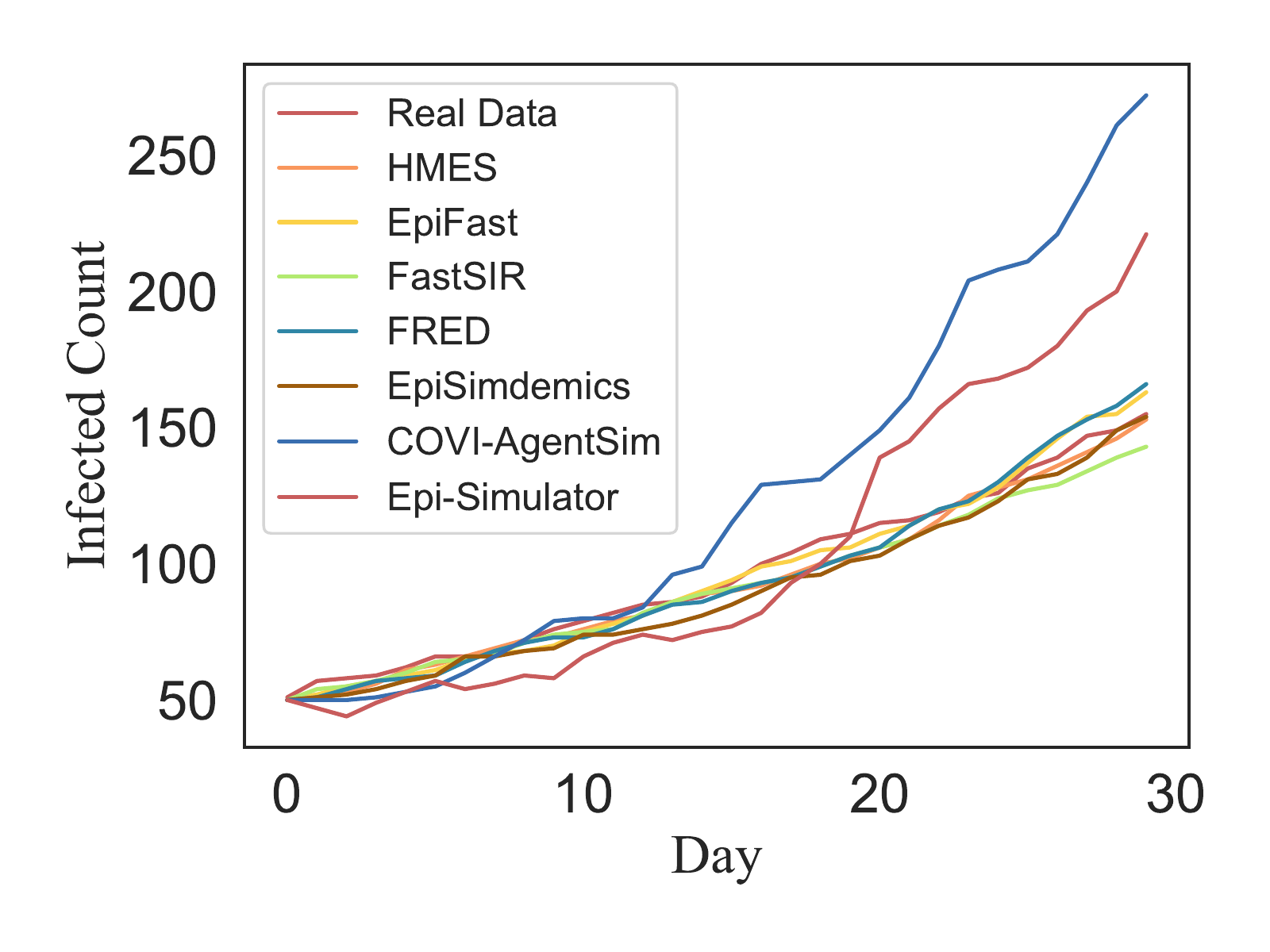}
        & \includegraphics[width=0.23\textwidth]{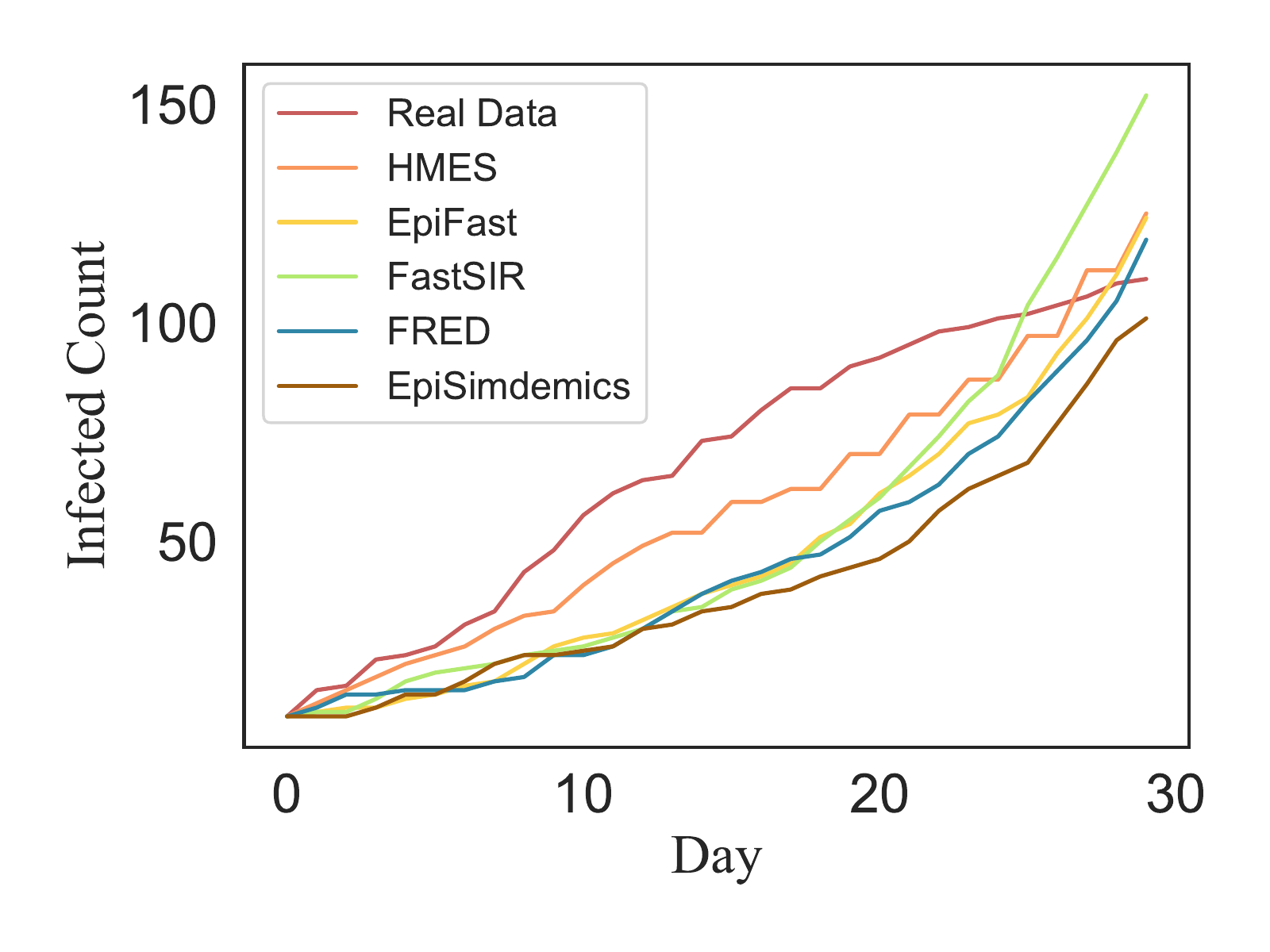}
        & \includegraphics[width=0.23\textwidth]{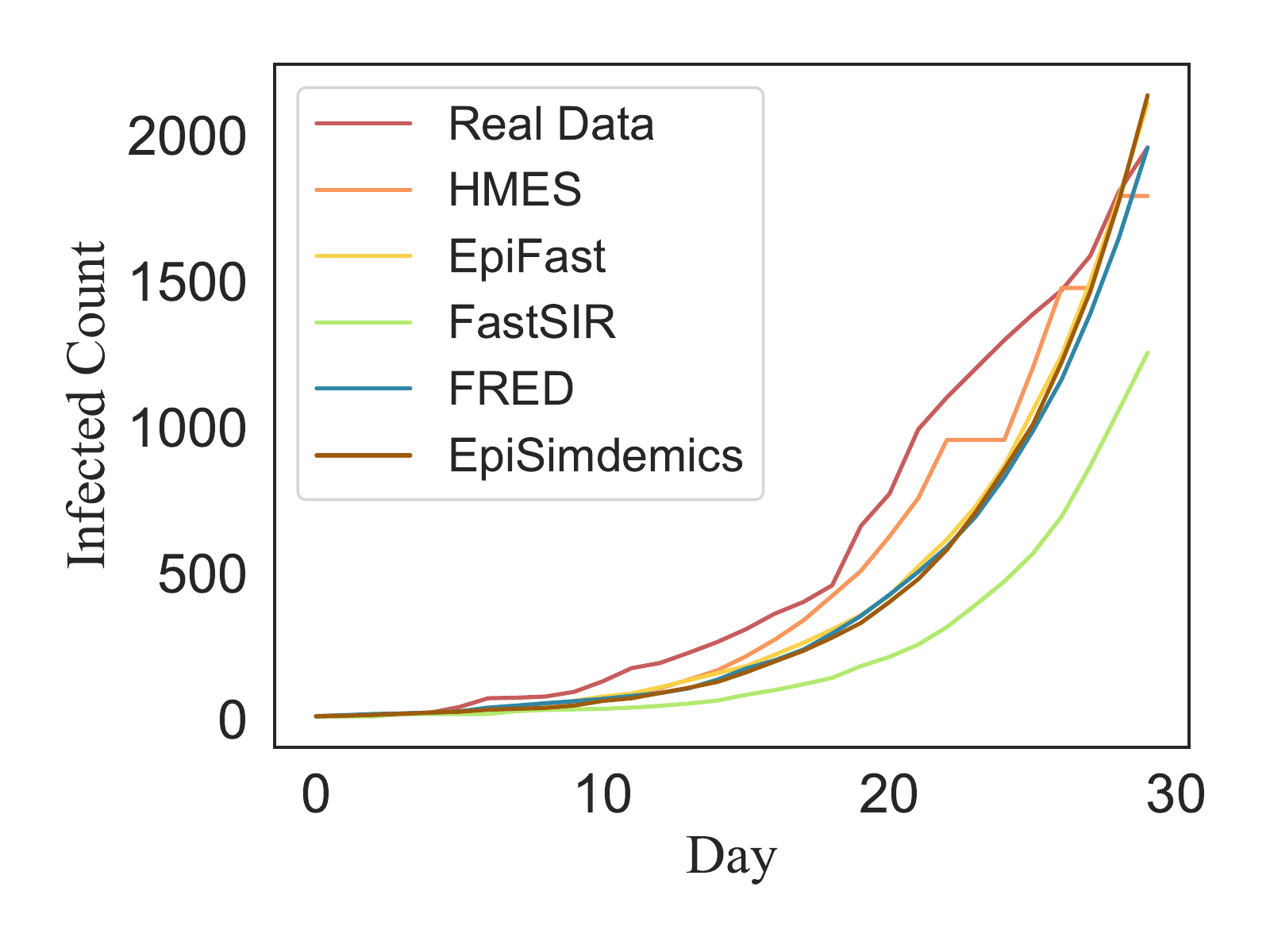}
        & \includegraphics[width=0.23\textwidth]{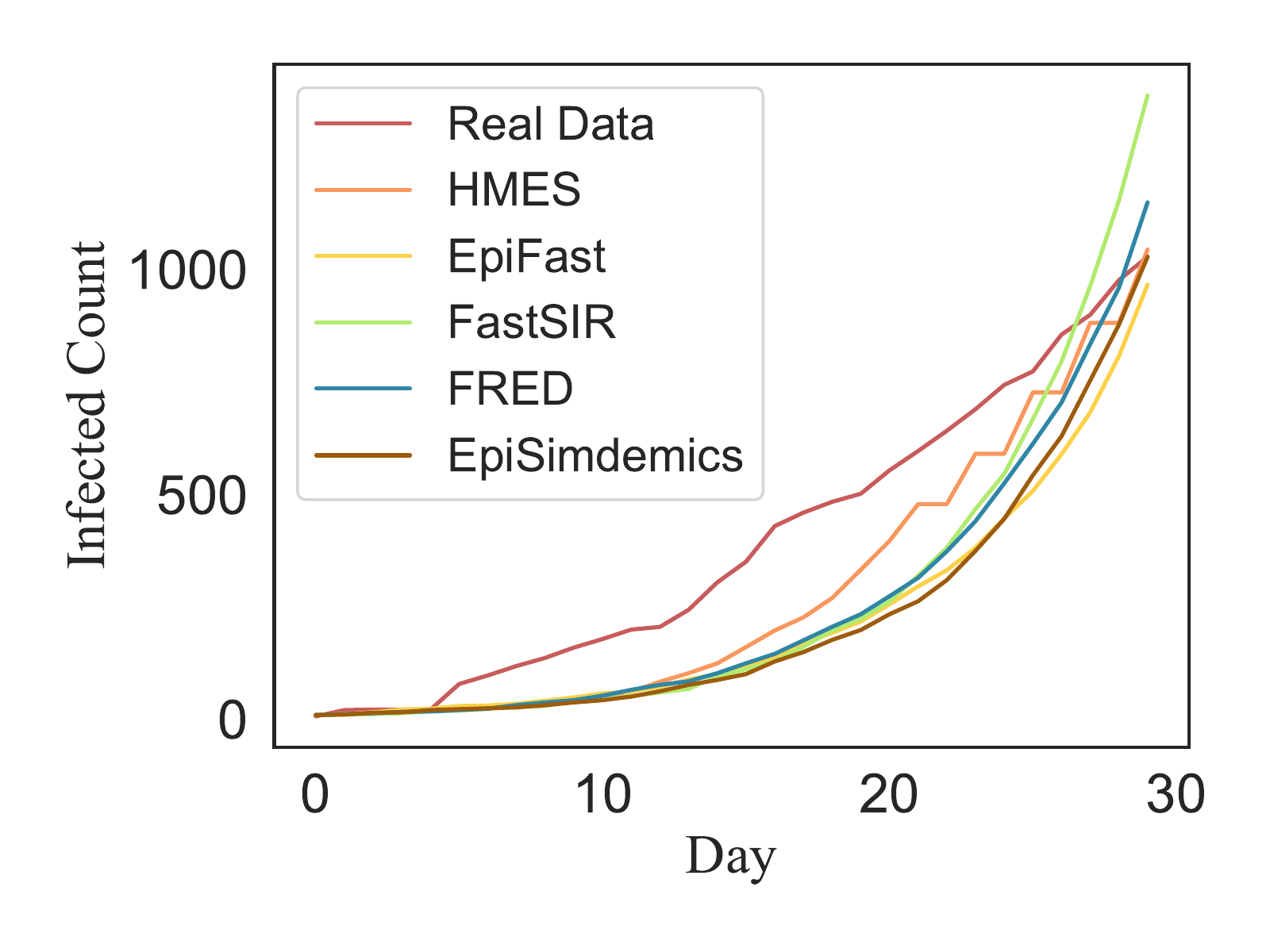} 
        \vspace{-0.25cm}
        \\
    (a) Small county & (b) Small city  & (c) Medium city &  (d) Large city
    \end{tabular}
    \caption{Infection curve for different methods for 4 example areas with various sizes in Table~\ref{tab:infection-case15}. \ours perform more accurate simulation not only on final count but also during the simulation process.}
    \label{fig:infection-curve}
    \vspace{-0.5cm}
\end{figure*}

\vspace{-3pt}
\subsection{Experiment Setup and Protocols}
\vspace{-5pt}
By default, each simulation runs for 30 simulation days ($D$), with each day containing 14 hours of activity ($H$) (the hours during which most people are inactive (9pm - 7am) are omitted). The initial infected count is set to be 10. 
For network-based methods, we ensure equivalent number of contacts per person to build the network. More on experimental settings are available in Appendix.

For simplicity, when describing population size and the number of locations, we will use `K' and `M' to represent thousands and millions respectively. All experiments are conducted on Intel Xeon (R) Gold 5118 @ 2.30GHz CPU. For efficiency comparison, we have ensured that different algorithms use the same number of cores.

{\bf Compared Methods}
We compare \ours with following state-of-the-art methods in epidemic prediction and control:
\begin{itemize}[leftmargin=*]
    \item \textbf{EpiSimdemics}~\cite{episimdemics} simulates large-scale epidemic progression, which computes infection by generating people's mobility and finding their shared trajectories.
    \item \textbf{FRED}~\cite{grefenstette2013fred} (A Framework for Reconstructing Epidemic Dynamics) is an epidemic modeling system that captures the demographic and geographic heterogeneities of the population and conducts pair-wise infection computation.
    \item \textbf{EpiFast}~\cite{epifast} computes pair-wise infection based on the stochastic disease propagation in the contact network.
    \item \textbf{FastSIR}~\cite{fastsir} is a state of the art method for network-based epidemic simulation that captures all infection transfers without epidemic dynamics in time.
    \item \textbf{Epi-Simulator}~\cite{hao2020understanding} builds epidemic simulator to estimate the efficacy of different mobility controls.
    \item \textbf{COVI-AgentSim}~\cite{bengio2020predicting} builds simulations and predict an individual’s infectiousness based on contact history.
\end{itemize}
Note that these methods are designed without intervention module originally. For a comprehensive comparison, we implement basic contact tracing algorithm as Alg.~\ref{alg:intervention-naive} (in Appendix) for all baselines. Among these, Epi-Simulator and COVI-AgentSim are compared in small scale as Figure~\ref{fig:intervention-curve} as the long running time on large scale data makes them unpractical.
We also note some recent works to predict and control the pedanmic falls into the categories for the baselines above\cite{feng2020learning,chang2021supporting,chang2021mobility}, and we merge them in experiments for simplicity.

\vspace{-5pt}
\subsection{Epidemic Progression Effectiveness}\label{sec:experi:effectiveness}
\vspace{-5pt}
\looseness=-1 To validate the simulation accuracy for the infection process, we conduct simulation on 15 cities with different sizes (divided by population) for 30 days. The population data and location data are collected from public census data\textsuperscript{\ref{note:census}}.
Some key statistics about the census data of various city sizes: five small cities of around 10K people and 25 tracts with average population density 3300, five medium cities of population ranging from 32K to 62K, with 90 tracts in average, and five large cities of over 1M or 2M people with around 300 tracts. 
Note that we fit one set of parameters for infection within cities of the same scale, afterwards adapt and generalize to others directly.

The simulation error on the 30th day is shown in Table~\ref{tab:infection-case15}, where \ours simulate the infection with lower error for majority of cases (12 out of 15). More importantly, the infection curve in Figure~\ref{fig:infection-curve} of \ours model lies much closer to the real data curve than others. This is due to the real-like mobility pattern and intervention strategies. 

\vspace{-5pt}
\subsection{Results on Efficiency}\label{sec:experi:efficiency}
\vspace{-5pt}
In this section, we compare the running time of different simulations. To make the comparison fair, the initial and final infected cases are set comparable for each method.

\begin{table}[tbph]
\vspace{-0.3cm}
\caption{Runtime comparison (in second) for 10M population}
\vspace{-0.3cm}
\label{tab:experi-10m}
\centering
\resizebox{0.45\textwidth}{!}{%
\begin{tabular}{c|c|c|c|c}
\toprule
Method       & \tabincell{c}{Mobility\\ Generation} & \tabincell{c}{Epidemic\\ Propagation} & \tabincell{c}{Intervention} & \tabincell{c}{Total   time} \\ 
\midrule
\textbf{\ours}         &\textbf{596.63}&	\textbf{709.62}&	\textbf{6.15}	&\textbf{1312.41}   \\ 
\tabincell{c}{Epi-\\Simdemics~\cite{episimdemics}} & 1057.16&	794.88&	1649.68	&3502.87    \\ 
FRED~\cite{grefenstette2013fred}         & 1058.23&	1023.28&	1632.83&	3715.36    \\ 
Epifast~\cite{epifast}      & 1486.81&	2020.56&	2425.21	&5932.96   \\ 
FastSIR~\cite{fastsir}      & 1501.52&	1351.54&	2642.36	&5496.14    \\ 
\bottomrule
\end{tabular}
}
\vspace{-0.1cm}
\end{table}

\looseness=-1 {\bf General comparison.} We compare the running time of different methods under 10M people and 10K locations setting. This has reached the similar scale of most metropolitan cities (Shanghai (20M), New York (8M)). The results are shown in Table~\ref{tab:experi-10m}. In general, \ours runs much faster in terms of the total time and time for all three modules. The biggest acceleration comes from the intervention module. The proposed fast contact tracing speeds up the intervention significantly. 


\begin{figure}[htbp]
  \vspace{-5mm}
  \centering 
  \subfigure[Total running time  ]{ 
    \includegraphics[width=0.22\textwidth]{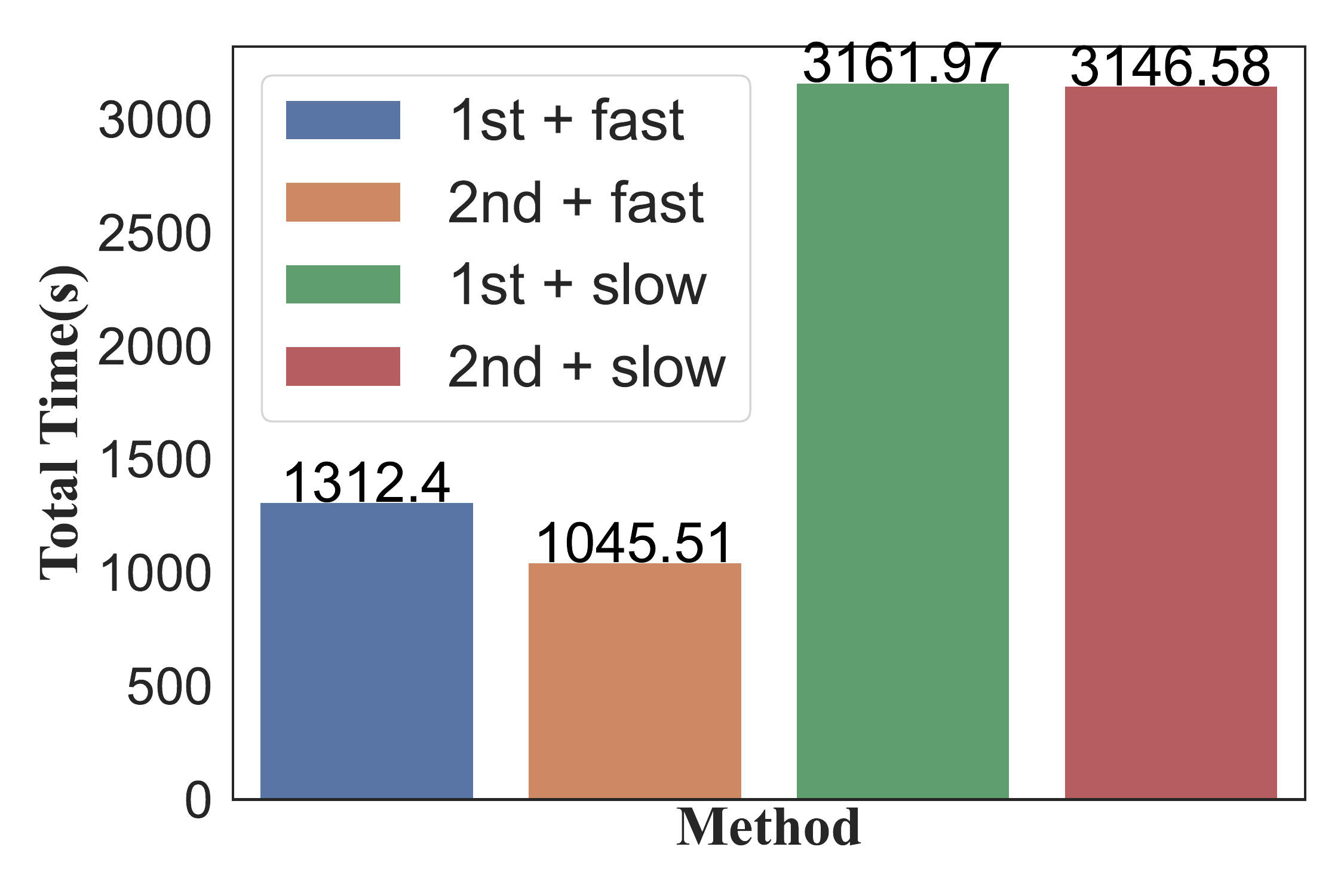}
  } 
  \hfill
  \subfigure[Runtime for intervention]{ 
    \includegraphics[width=0.22\textwidth]{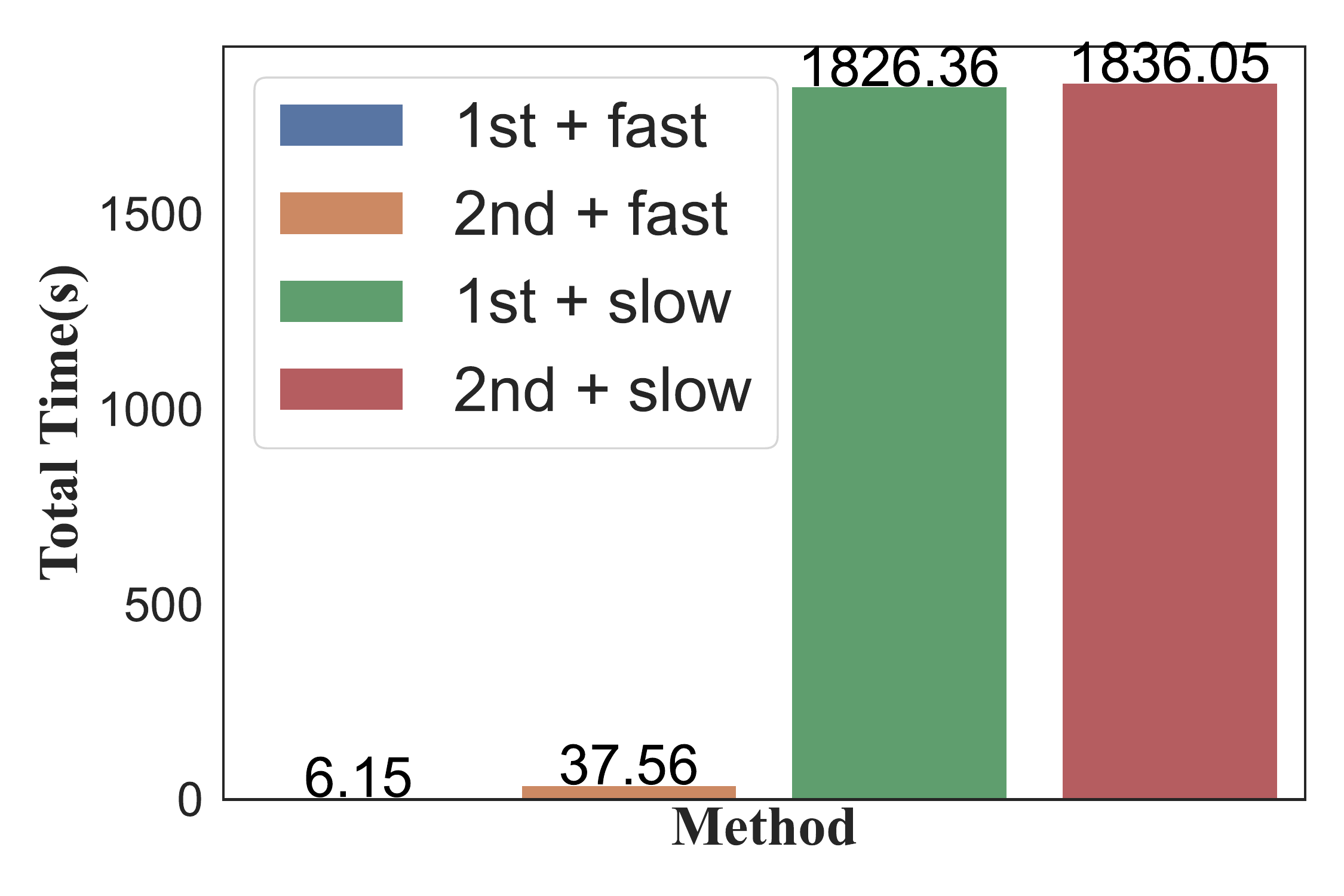}
  } 
    \vspace{-4.5mm}
  \caption{Runtime for intervention. `Fast', `Slow' for Alg.\ref{alg:intervention}, for Alg. \ref{alg:intervention-naive}} 
\label{fig:intervention}
  \vspace{-0.8mm}
\end{figure}




Then we test the running time of all compared methods under different parameters, including different intervention strategies, and more for population size, population density and infection rate elaborated in Appendix. 

\looseness=-1 {\bf Intervention Strategy.} We conduct experiments to test the acceleration of our fast contact tracing under different intervention strategies in a  1M-population and 1K=location city. The default intervention strategy is to hospitalize infected people and isolate their contacts. We perform first-order and second-order tracing using Alg.~2(in main paper) 
(denoted as `fast') and basic contact tracing (denoted as `slow'). The total running time and running time for the intervention module are shown in Figure~\ref{fig:intervention}(a) and Figure~\ref{fig:intervention}(b) respectively. As is shown, \ours  accelerates around 300 times and 50 times for 1st-order and 2nd-order, respectively. Note that the total time (as shown in Figure~\ref{fig:intervention}(a)) is smaller in second-order because isolating second-order contacts decreases the number of people who can move freely, hence reduces the running time.

\vspace{-5pt}
\subsection{Effects of Different Intervention Strategies}\label{sec:experi:intervene-effectiveness}
\vspace{-6pt}
We evaluate the effectiveness and necessity of individual-level intervention compared to the previous coarse-grained group-level intervention\cite{chang2021supporting,chang2021mobility,feng2020learning}.
As a simple case, the experiment setting is on 10K people and 100 locations with infection rate 0.05. The incubation step is 56 (four days), and the mobility randomness is 0.5. The intervention strategy is to isolate the given people group for 5 days.

\begin{figure*}[tbhp]
    \centering
    \begin{tabular}{ccc}
        \includegraphics[width=0.25\textwidth]{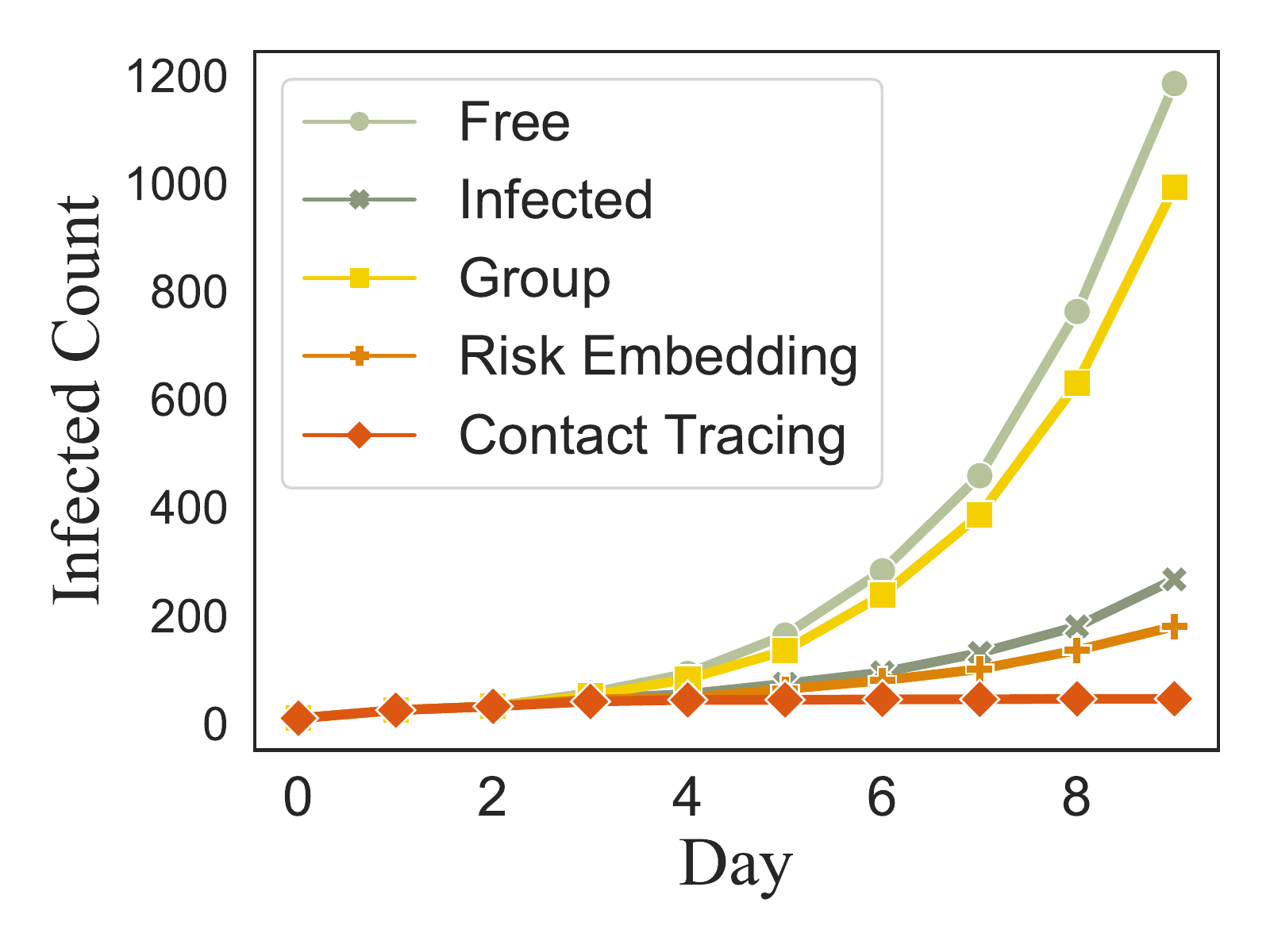}
        & \includegraphics[width=0.25\textwidth]{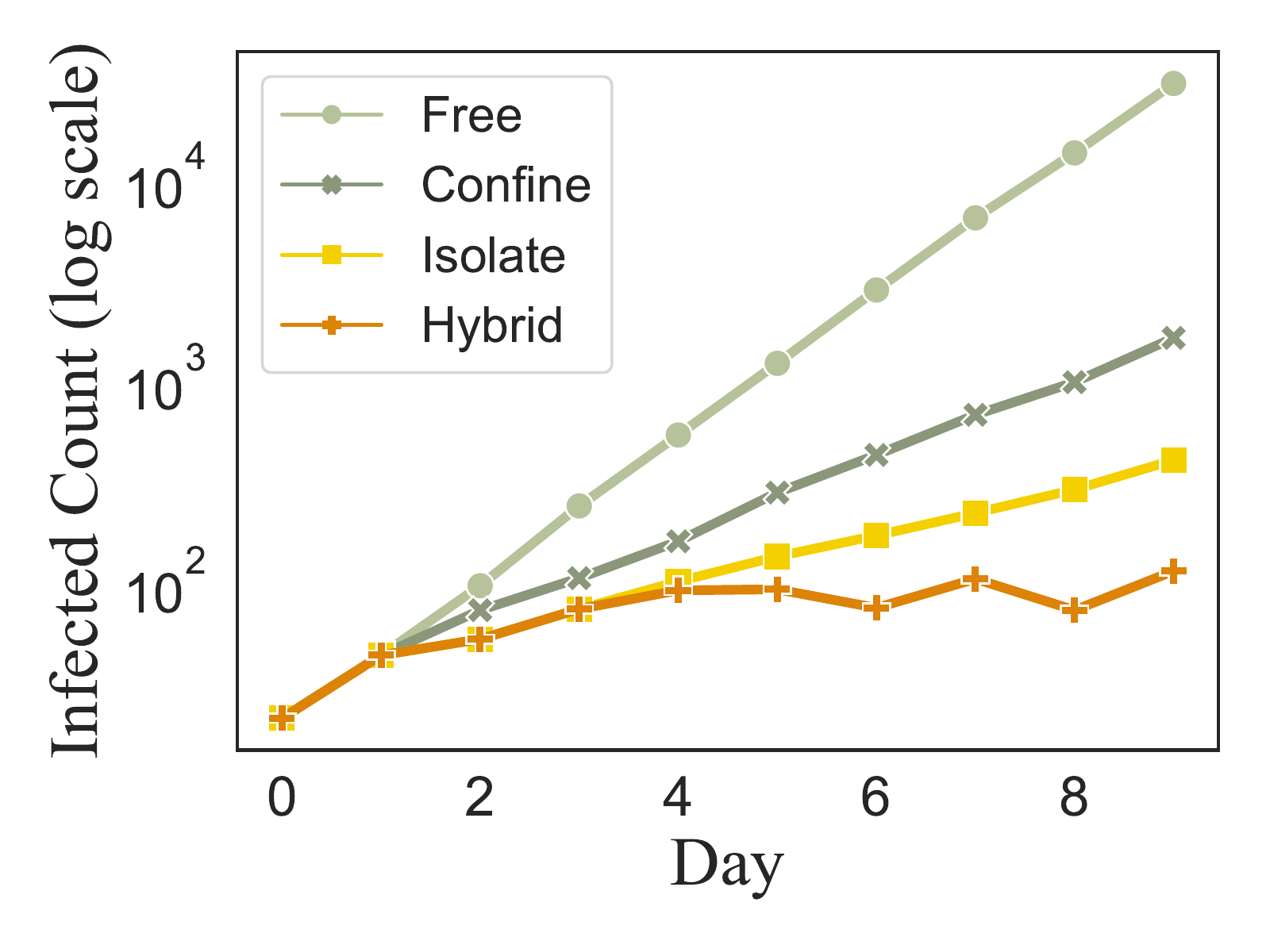}
        & \includegraphics[width=0.25\textwidth]{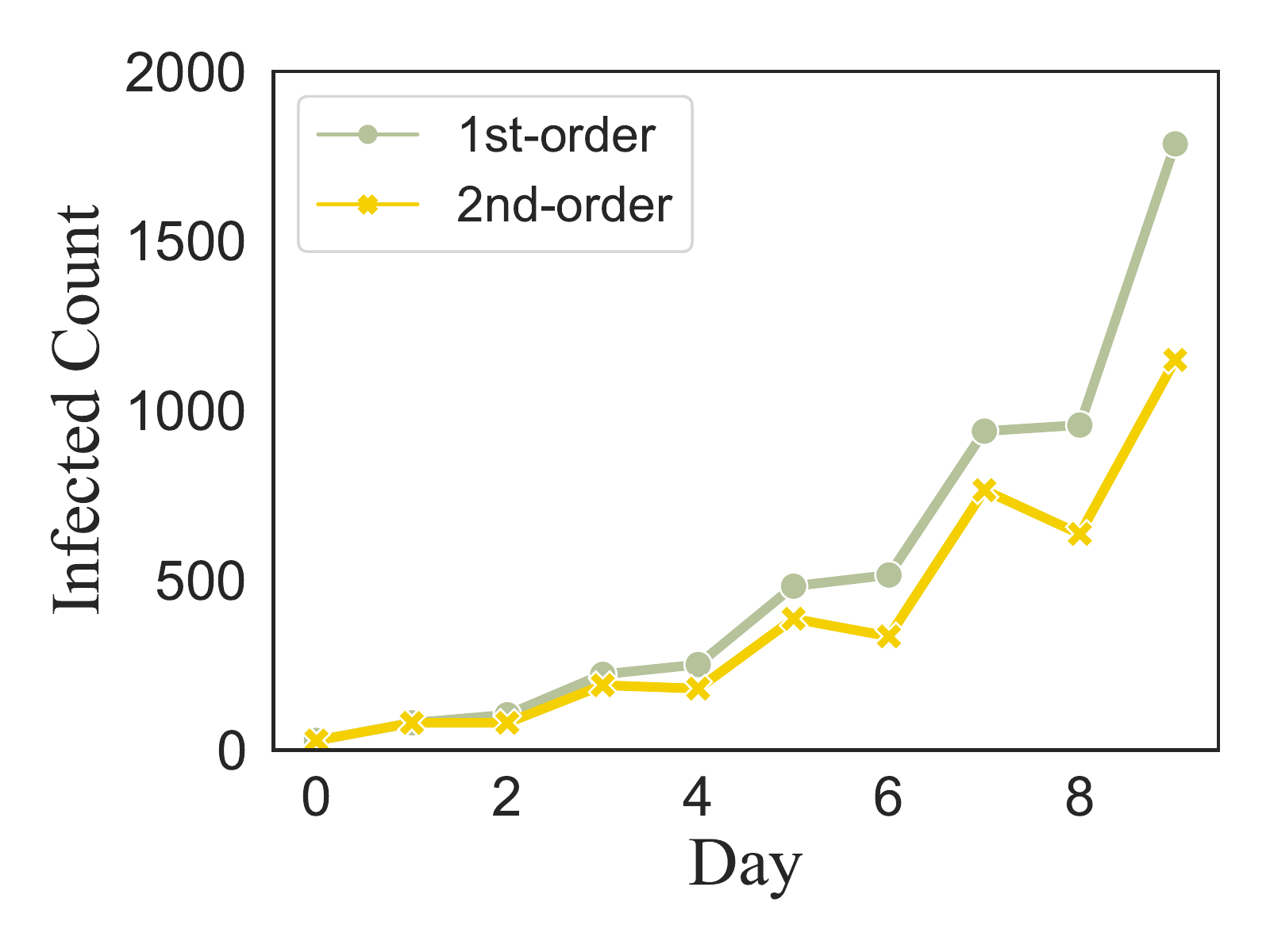}      
        \vspace{-0.3cm}
        \\
        (a) Individual v.s. group &
        (b) Different strategies   & (c) Order of tracing \\
    \end{tabular}
    \vspace{-0.2cm}
    \caption{Infection curve for various intervention settings}
    \label{fig:intervention-curve}
    \vspace{-0.7cm}
\end{figure*}

\looseness=-1 {\bf Individual-level intervention v.s. group-level intervention} We first compare individual-level interventions with group-level ones on a scenario with 100 locations and 10K population. Concretely, we compare intervention results of the following three categories of methods, (1) No intervention (Free); (2) Group-level control (Group); and (3) Individual-level control (Infected, Risk Embedding(shown in Appendix), Contact Tracing), where "Infected" is to isolate all infected people. All the methods are given the same quota of intervention resources (i.e., the number of isolated people) except the contact tracing method. The results are shown in Figure~\ref{fig:intervention-curve}(a). We can observe that individual-level policies generally perform better than group-level policies and no intervention. In addition, only constraining the confirmed cases will not stop the pandemic, due to the existence of incubation periods. Contact tracing method can mitigate the pandemic to the most extent although it requires more medical resources.

{\bf Intervention Intensity} To verify the effectiveness of different intervention intensities (free, confine, isolate, hospitalize), we implement these strategies on a scenario with 100 locations and 10K population. For confine and isolate, we set restrictions on both the infected and their contacts. For the hybrid strategy, we hospitalize the infected while isolate their contacts. As shown in Figure~\ref{fig:intervention-curve}(b), the `free', `confine' and `isolate' strategy all lead to linear curve in the log scale. Equivalently, we have the following observation. Without any restriction (`free'), the infected count increases exponentially with time. Using `confine' and `isolate' strategy helps slow down the increasing trend, but the curve is still exponential. When the `hybrid' strategy is applied, the curve flattens in several days. 

{\bf Multi-order Tracing} Furthermore, we test the influence of 1st- and 2nd-order contact tracing by showing the infected count in Figure~\ref{fig:intervention-curve}(c). As expected, the latter further suppresses the spread of epidemic compared to the former.
\nop{
Due to the existence of incubation periods, intervention on the confirmed cases only does not block the propagation of pandemic. we expect our algorithm to discover and intervene further on people that are under potential infection risk without cumbersome contact tracing. Figure~\ref{Fig.methodology} illustrates the case that our algorithm could do better than direct symptomatic intervention, while the baseline is to intervene a random number besides the infected. Under exactly the same quota of interventions (the same social resource consumption), our supported individual-based performs the least number of cases. Under lower than five percent of intervention quotas, the intervention based on a simple prediction could discover the risky people under incubation. Our simulation supports more abundant and complex policies with a portable interface.
}
\vspace{-5pt}
\subsection{Demonstrations and Case Studys}\label{sec:case study}
\vspace{-5pt}
{\bf Python Interfaces}
Upon implementation of C language, an interface is provided for mobility, infection and intervention simulations in Python package. The package provide basic function calls where users could customize any simulation. It has also been used by Prescriptive Analytics for the Physical World (PAPW)~\footnote{https://prescriptive-analytics.github.io/\label{note:papw}} Challenge for pandemic mobile intervention competition.
\begin{figure}[H]
    \vspace{-0.45cm}
    \centering
    \includegraphics[width=0.43\textwidth]{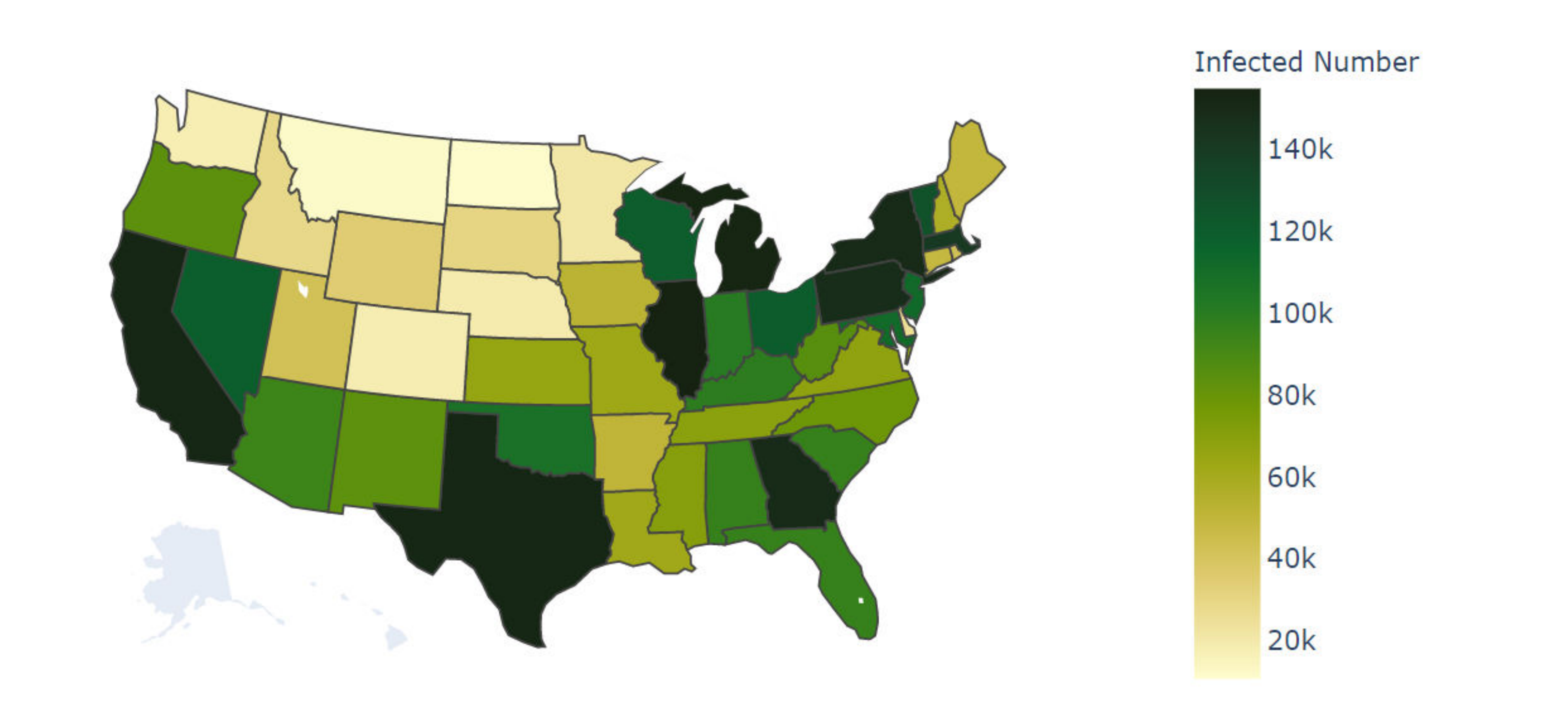}
    \vspace{-0.45cm}
    \caption{Simulated Epidemic Situation 180 Days after Outbreak}
    \label{fig:case study}
    \vspace{-0.5cm}
\end{figure}
{\bf Case Study on Country-wise Progression}
To evaluate on large-scale data, we use the US Census data~\footnote{https://www.census.gov\label{note:census}} to conduct a simulation of the whole United States (Alaska and Hawaii are not included) for 180 days. As shown in Figure~\ref{fig:case study}, the epidemic spread across country, with several densely populated states having more than 100K people infected. The infected count in \ours after 180 days is 3,887,528 compared to 3,957,593 in real data  with the relative error 1.77\%. Details of Python interface and specific settings here are shown in Appendix.

\nop{
In Figure~\ref{fig:comparison-HMS-real case}(d), we impose a typical intervention strategy on the large city simulation. The intervention is to hospitalize infected people and isolate their first-order contacts. As seen in the curve without intervention, there is a breakout on day 8 followed by exponential growth on day 9 and 10. However, it is  much flatter and almost turns to be linear after adopting intervention measures, which demonstrates the effectiveness of our intervention strategy in a trial experiment.}
\vspace{-3pt}
\section{Conclusion}\label{sec:conclusion}
\vspace{-5pt}
In this paper, we propose a scalable simulation framework \ours for epidemic simulation and control. Compared to previous ones, we provide a large-scale simulation platform supporting flexible intervention interfaces on individual and group basis. 
We conduct extensive experiments on evaluation of accuracy, efficiency and flexibility.
We hope \ours could boost future researches in human mobility analysis, and help develop tools for more complex, effective and humanized policies to mitigate pandemics in real world.
\vspace{-5pt}
\section{Acknowledgement}
\vspace{-6pt}
This work was sponsored by Shanghai Pujiang Program under Grant No. 20PJ1409400, National Natural Science Foundation of China under Grant No. 62102246 and Provincial Key Research and Development Program of Zhejiang under Grant No. 2021C01034. This work was done in collaboration with Hangzhou Yunqi Academy of Engineering.

{
\bibliographystyle{abbrv}
\looseness=-1
\bibliography{ref_cut}

\begin{thebibliography}{10}

\bibitem{kdd19-epideep}
B.~Adhikari, X.~Xu, N.~Ramakrishnan, and B.~A. Prakash.
\newblock Epideep: Exploiting embeddings for epidemic forecasting.
\newblock In {\em SIGKDD}, 2019.

\bibitem{fastsir}
N.~Antulov-Fantulin, A.~Lan{\v{c}}i{\'c}, H.~{\v{S}}tefan{\v{c}}i{\'c}, and
  M.~{\v{S}}iki{\'c}.
\newblock Fastsir algorithm: A fast algorithm for the simulation of the
  epidemic spread in large networks by using the
  susceptible--infected--recovered compartment model.
\newblock {\em Information sciences}, 239:226--240, 2013.

\bibitem{arenas2020mathematical}
A.~Arenas, W.~Cota, J.~G{\'o}mez-Gardenes, S.~G{\'o}mez, C.~Granell, J.~T.
  Matamalas, D.~Soriano-Panos, and B.~Steinegger.
\newblock A mathematical model for the spatiotemporal epidemic spreading of
  covid19.
\newblock 2020.

\bibitem{arenas2020modeling}
A.~Arenas, W.~Cota, J.~G{\'o}mez-Garde{\~n}es, S.~G{\'o}mez, C.~Granell, J.~T.
  Matamalas, D.~Soriano-Pa{\~n}os, and B.~Steinegger.
\newblock Modeling the spatiotemporal epidemic spreading of covid-19 and the
  impact of mobility and social distancing interventions.
\newblock {\em Physical Review X}, 2020.

\bibitem{2010sirs}
J.-D. Bancal and R.~Pastor-Satorras.
\newblock Steady-state dynamics of the forest fire model on complex networks.
\newblock {\em The European Physical Journal}, 2010.

\bibitem{episimdemics}
C.~L. {Barrett}, K.~R. {Bisset}, S.~G. {Eubank}, {Xizhou Feng}, and M.~V.
  {Marathe}.
\newblock Episimdemics: An efficient algorithm for simulating the spread of
  infectious disease over large realistic social networks.
\newblock In {\em ACM/IEEE Conference on Supercomputing}, pages 1--12, 2008.

\bibitem{bengio2020predicting}
Y.~Bengio, P.~Gupta, T.~Maharaj, N.~Rahaman, M.~Weiss, T.~Deleu, E.~B. Muller,
  M.~Qu, P.-l. St-charles, O.~Bilaniuk, et~al.
\newblock Predicting infectiousness for proactive contact tracing.
\newblock In {\em International Conference on Learning Representations}, 2020.

\bibitem{bhatele2017massively}
A.~Bhatele, J.-S. Yeom, N.~Jain, C.~J. Kuhlman, Y.~Livnat, K.~R. Bisset, L.~V.
  Kale, and M.~V. Marathe.
\newblock Massively parallel simulations of spread of infectious diseases over
  realistic social networks.
\newblock In {\em IEEE/ACM International Symposium on Cluster, Cloud and Grid
  Computing}, 2017.

\bibitem{epifast}
K.~R. Bisset, J.~Chen, X.~Feng, V.~A. Kumar, and M.~V. Marathe.
\newblock Epifast: A fast algorithm for large scale realistic epidemic
  simulations on distributed memory systems.
\newblock In {\em International Conference on Supercomputing}, ICS ’09, page
  430–439. ACM, 2009.

\bibitem{bisset2009modeling}
K.~R. Bisset, X.~Feng, M.~Marathe, and S.~Yardi.
\newblock Modeling interaction between individuals, social networks and public
  policy to support public health epidemiology.
\newblock In {\em the 2009 Winter Simulation Conference (WSC)}, pages
  2020--2031. IEEE, 2009.

\bibitem{biswas2014seir}
M.~H.~A. Biswas, L.~T. Paiva, and M.~De~Pinho.
\newblock A seir model for control of infectious diseases with constraints.
\newblock {\em Mathematical Biosciences \& Engineering}, 11(4):761, 2014.

\bibitem{brockmann2006scaling}
D.~Brockmann, L.~Hufnagel, and T.~Geisel.
\newblock The scaling laws of human travel.
\newblock {\em Nature}, 439(7075):462--465, 2006.

\bibitem{browne2021infection}
C.~A. Browne, D.~B. Amchin, J.~Schneider, and S.~S. Datta.
\newblock Infection percolation: A dynamic network model of disease spreading.
\newblock {\em Frontiers in Physics}, 9:645954, 2021.

\bibitem{chang2021mobility}
S.~Chang, E.~Pierson, P.~W. Koh, J.~Gerardin, B.~Redbird, D.~Grusky, and
  J.~Leskovec.
\newblock Mobility network models of covid-19 explain inequities and inform
  reopening.
\newblock {\em Nature}, 589(7840):82--87, 2021.

\bibitem{chang2021supporting}
S.~Y. Chang, M.~L. Wilson, B.~Lewis, Z.~Mehrab, K.~K. Dudakiya, E.~Pierson,
  P.~W. Koh, J.~Gerardin, B.~Redbird, D.~Grusky, et~al.
\newblock Supporting covid-19 policy response with large-scale mobility-based
  modeling.
\newblock {\em medRxiv}, 2021.

\bibitem{chinazzi2020effect}
M.~Chinazzi, J.~T. Davis, M.~Ajelli, C.~Gioannini, M.~Litvinova, S.~Merler,
  A.~P. y~Piontti, K.~Mu, L.~Rossi, K.~Sun, et~al.
\newblock The effect of travel restrictions on the spread of the 2019 novel
  coronavirus (covid-19) outbreak.
\newblock {\em Science}, 368(6489):395--400, 2020.

\bibitem{episims}
S.~Eubank, H.~Guclu, V.~A. Kumar, M.~V. Marathe, A.~Srinivasan, Z.~Toroczkai,
  and N.~Wang.
\newblock Modelling disease outbreaks in realistic urban social networks.
\newblock {\em Nature}, 429(6988):180--184, 2004.

\bibitem{feng2020learning}
J.~Feng, Z.~Yang, F.~Xu, H.~Yu, M.~Wang, and Y.~Li.
\newblock Learning to simulate human mobility.
\newblock In {\em ACM SIGKDD}, 2020.

\bibitem{ferguson2020report}
N.~Ferguson, D.~Laydon, G.~Nedjati~Gilani, N.~Imai, K.~Ainslie, M.~Baguelin,
  S.~Bhatia, A.~Boonyasiri, Z.~Cucunuba~Perez, G.~Cuomo-Dannenburg, et~al.
\newblock Report 9: Impact of non-pharmaceutical interventions (npis) to reduce
  covid19 mortality and healthcare demand.
\newblock 2020.

\bibitem{flaxman2020report}
S.~Flaxman, S.~Mishra, A.~Gandy, H.~Unwin, H.~Coupland, T.~Mellan, H.~Zhu,
  T.~Berah, J.~Eaton, P.~Perez~Guzman, et~al.
\newblock Report 13: Estimating the number of infections and the impact of
  non-pharmaceutical interventions on covid-19 in 11 european countries.
\newblock 2020.

\bibitem{flaxman2020estimating}
S.~Flaxman, S.~Mishra, A.~Gandy, H.~J.~T. Unwin, T.~A. Mellan, H.~Coupland,
  C.~Whittaker, H.~Zhu, T.~Berah, J.~W. Eaton, et~al.
\newblock Estimating the effects of non-pharmaceutical interventions on
  covid-19 in europe.
\newblock {\em Nature}, 584(7820):257--261, 2020.

\bibitem{gonzalez2008understanding}
M.~C. Gonzalez, C.~A. Hidalgo, and A.-L. Barabasi.
\newblock Understanding individual human mobility patterns.
\newblock {\em nature}, 453(7196):779--782, 2008.

\bibitem{grefenstette2013fred}
J.~J. Grefenstette, S.~T. Brown, R.~Rosenfeld, J.~DePasse, N.~T. Stone, P.~C.
  Cooley, W.~D. Wheaton, A.~Fyshe, D.~D. Galloway, A.~Sriram, et~al.
\newblock Fred (a framework for reconstructing epidemic dynamics):an
  open-source software system for modeling infectious diseases and control
  strategies using census-based populations.
\newblock {\em BMC public health}, 2013.

\bibitem{gupta2020covi}
P.~Gupta, T.~Maharaj, M.~Weiss, N.~Rahaman, H.~Alsdurf, A.~Sharma, N.~Minoyan,
  S.~Harnois-Leblanc, V.~Schmidt, P.-L.~S. Charles, et~al.
\newblock Covi-agentsim: an agent-based model for evaluating methods of digital
  contact tracing.
\newblock {\em arXiv}, 2020.

\bibitem{hao2020understanding}
Q.~Hao, L.~Chen, F.~Xu, and Y.~Li.
\newblock Understanding the urban pandemic spreading of covid-19 with real
  world mobility data.
\newblock In {\em the 26th ACM SIGKDD}, pages 3485--3492, 2020.

\bibitem{hsiang2020effect}
S.~Hsiang, D.~Allen, S.~Annan-Phan, K.~Bell, I.~Bolliger, T.~Chong,
  H.~Druckenmiller, L.~Y. Huang, A.~Hultgren, E.~Krasovich, et~al.
\newblock The effect of large-scale anti-contagion policies on the covid-19
  pandemic.
\newblock {\em Nature}, 584(7820):262--267, 2020.

\bibitem{1927SIR}
W.~O. Kermack and A.~G. McKendrick.
\newblock A contribution to the mathematical theory of epidemics.
\newblock {\em the royal society of london. Series A, Containing papers of a
  mathematical and physical character}, 1927.

\bibitem{EpidemicSim}
S.~{Kopman}, M.~I. {Akbaş}, and D.~{Turgut}.
\newblock Epidemicsim: Epidemic simulation system with realistic mobility.
\newblock In {\em IEEE Conference on Local Computer Networks - Workshops},
  pages 659--665, 2012.

\bibitem{kraemer2020effect}
M.~U. Kraemer, C.-H. Yang, B.~Gutierrez, C.-H. Wu, B.~Klein, D.~M. Pigott,
  L.~Du~Plessis, N.~R. Faria, R.~Li, W.~P. Hanage, et~al.
\newblock The effect of human mobility and control measures on the covid-19
  epidemic in china.
\newblock {\em Science}, 368(6490):493--497, 2020.

\bibitem{kuhn1955hungarian}
H.~W. Kuhn.
\newblock The hungarian method for the assignment problem.
\newblock {\em Naval research logistics quarterly}, 2(1-2):83--97, 1955.

\bibitem{li2020early}
Q.~Li, X.~Guan, P.~Wu, X.~Wang, L.~Zhou, Y.~Tong, R.~Ren, K.~S. Leung, E.~H.
  Lau, J.~Y. Wong, et~al.
\newblock Early transmission dynamics in wuhan, china, of novel
  coronavirus--infected pneumonia.
\newblock {\em New England journal of medicine}, 2020.

\bibitem{lorch2020quantifying}
L.~Lorch, H.~Kremer, W.~Trouleau, S.~Tsirtsis, A.~Szanto, B.~Sch{\"o}lkopf, and
  M.~Gomez-Rodriguez.
\newblock Quantifying the effects of contact tracing, testing, and containment
  measures in the presence of infection hotspots.
\newblock 2020.

\bibitem{luo2020deeptrack}
Y.~Luo, W.~Li, T.~Zhao, X.~Yu, L.~Zhang, G.~Li, and N.~Tang.
\newblock Deeptrack: Monitoring and exploring spatio-temporal data: a case of
  tracking covid-19.
\newblock {\em the VLDB Endowment}, 13(12):2841--2844, 2020.

\bibitem{mcinerney2013breaking}
J.~McInerney, S.~Stein, A.~Rogers, and N.~R. Jennings.
\newblock Breaking the habit: Measuring and predicting departures from routine
  in individual human mobility.
\newblock {\em Pervasive and Mobile Computing}, 9(6):808--822, 2013.

\bibitem{nussbaumer2020quarantine}
B.~Nussbaumer-Streit, V.~Mayr, A.~I. Dobrescu, A.~Chapman, E.~Persad,
  I.~Klerings, G.~Wagner, U.~Siebert, D.~Ledinger, C.~Zachariah, et~al.
\newblock Quarantine alone or in combination with other public health measures
  to control covid-19: a rapid review.
\newblock {\em Cochrane Database of Systematic Reviews}, (9), 2020.

\bibitem{pappalardo2018data}
L.~Pappalardo and F.~Simini.
\newblock Data-driven generation of spatio-temporal routines in human mobility.
\newblock {\em Data Mining\&Knowledge Discovery}, 2018.

\bibitem{song2020reinforced}
S.~Song, Z.~Zong, Y.~Li, X.~Liu, and Y.~Yu.
\newblock Reinforced epidemic control: Saving both lives and economy.
\newblock {\em arXiv}, 2020.

\bibitem{stanley2018many}
K.~Stanley, E.-H. Yoo, T.~Paul, and S.~Bell.
\newblock How many days are enough?: capturing routine human mobility.
\newblock {\em International Journal of Geographical Information Science},
  32(7):1485--1504, 2018.

\bibitem{kdd2018-intracity}
J.~Wang, X.~Wang, and J.~Wu.
\newblock Inferring metapopulation propagation network for intra-city epidemic
  control and prevention.
\newblock In {\em ACM SIGKDD}, 2018.

\bibitem{xu2019stand}
F.~Xu, B.~Desmarais, and D.~Peuquet.
\newblock Stand: A spatio-temporal algorithm for network diffusion simulation.
\newblock {\em arXiv}, 2019.

\end{thebibliography}
}
\appendix
\section{Algorithm Details}

	\subsection{Epidemic Propogation}
	We provide the detailed epidemic progression as in Alg.\ref{alg:propagation}.
	\begin{algorithm}[ht]  
		\caption{Epidemic Progression}  
		\label{alg:propagation}  
		\KwIn{Current step $t$, people set $\mathbf{M}$, location set $\mathbf{L}$, people set that visited locations $\mathbf{V}$}
		\KwOut{Newly infected people $\mathbf{F}^{t}$ for step $t$} 
		\For{$l \in L$}{
			Update people set $\mathbf{I}_l^t$, $\mathbf{S}_l^t$, $\mathbf{R}_l^t$ of for location $l$ at time $t$ \\
			Calculate infection rate $p_l = p \cdot \frac{|\mathbf{I}_l^t|}{|\mathbf{S}_l^t| +|\mathbf{I}_l^t|+|\mathbf{R}_l^t|}$\\
			Infer newly infected count $|\mathbf{F}_l^t|= p_l \cdot |\mathbf{S}_l^t$|\\
			Randomly select $k_l$ people from $\mathbf{S}_l^t$ to form  the newly infected $\mathbf{F}_l^t$ \\
		}
	\end{algorithm}

	\subsection{Intevention}
	\subsubsection{Basic Contact Tracing}
	We demonstrate the basic contact tracing as in Alg.~\ref{alg:intervention-naive}.
	Concretely, the separation for each individual  has four levels, free (no restriction), confine (restricted at residential area), 
	isolate (can not get in touch with anyone), and hospitalize. This individual-level separation strategy support would enable further researches on more effective and complex intervention policies.
	


	\subsubsection{Improved Contact Tracing with bipartite check-tree}
	
	{\bf Complexity Analysis}
	We compare the time complexity of first order contact tracing, where that for higher order is similar. Note that the time for initialization and imposing intervention $\beta$ is omitted because they are neglectable with that of contact tracing. Since the original contact tracing algorithm loops over potential infectious people set $\mathbf{B}$, tracing steps $\tau$, and people co-locate with each queried person in each time step, the average time complexity should be $O(|\mathbf{B}| \cdot \tau \cdot  q)$. 
	The time complexity of that in Alg.2(in main paper)
	is $O(|\mathbf{B}| \cdot \tau + |\mathbf{C}_L| \cdot q) $. 
	Due to wide existence of people's co-occurrence in regular patterns of mobility, $|\mathbf{C}^\star_L| << |\mathbf{B}| \cdot \tau$, Alg.2(in main paper)
	is much faster than the original Alg.\ref{alg:intervention-naive}.

	\begin{figure*}[t!]
		\centering
		\begin{tabular}{ccc}
			\includegraphics[width=0.25\textwidth]{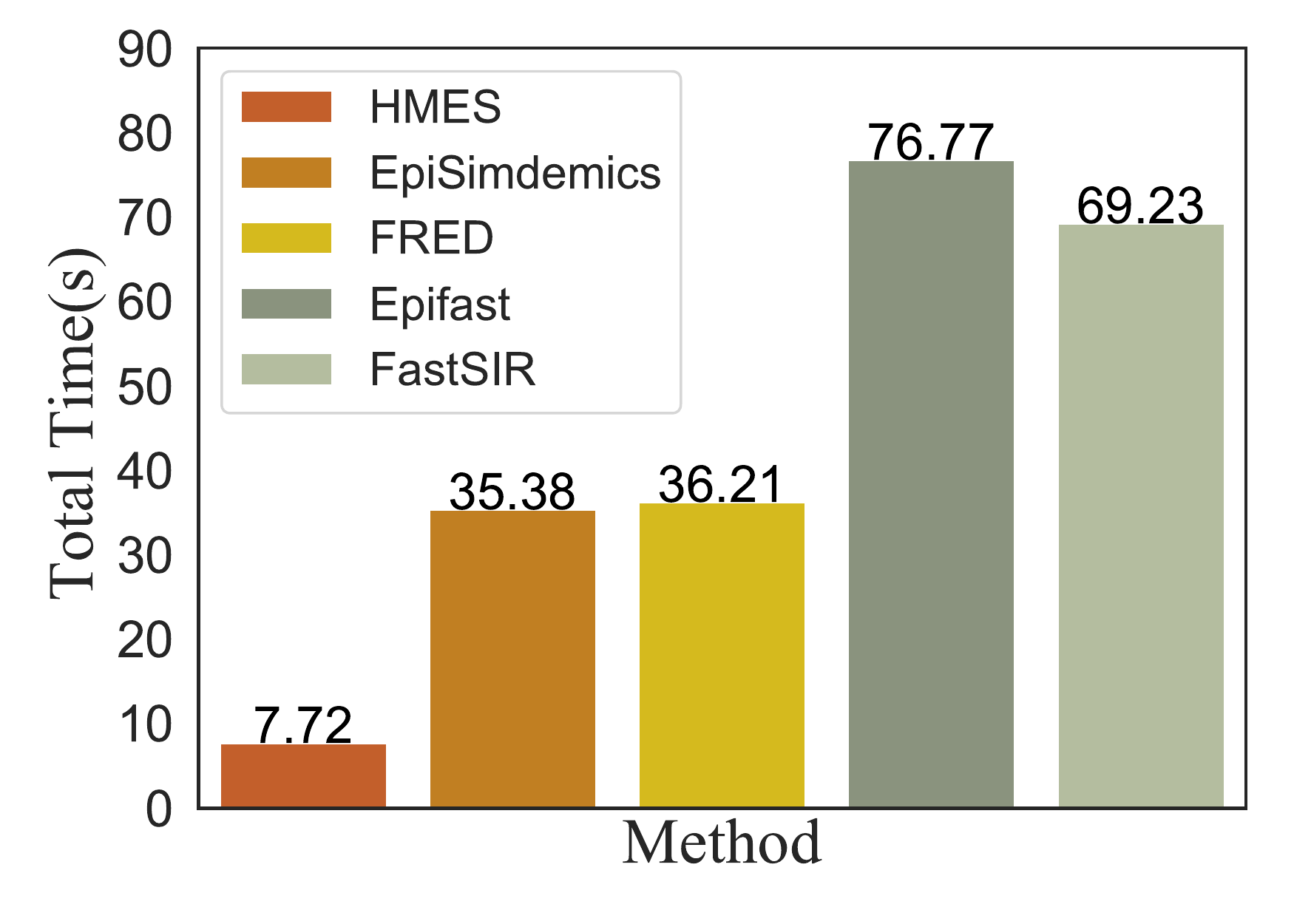}
			& \includegraphics[width=0.25\textwidth]{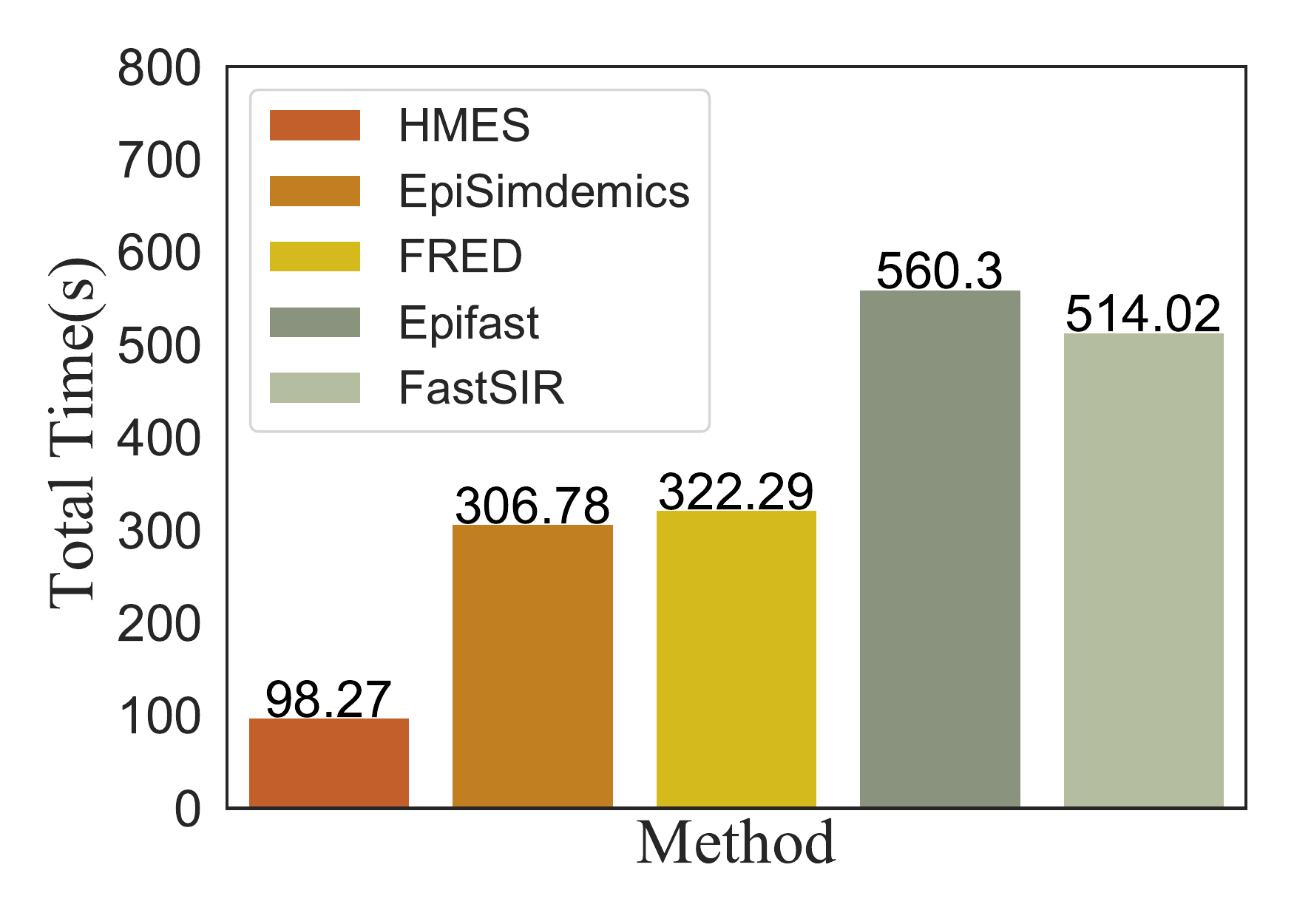}
			& \includegraphics[width=0.25\textwidth]{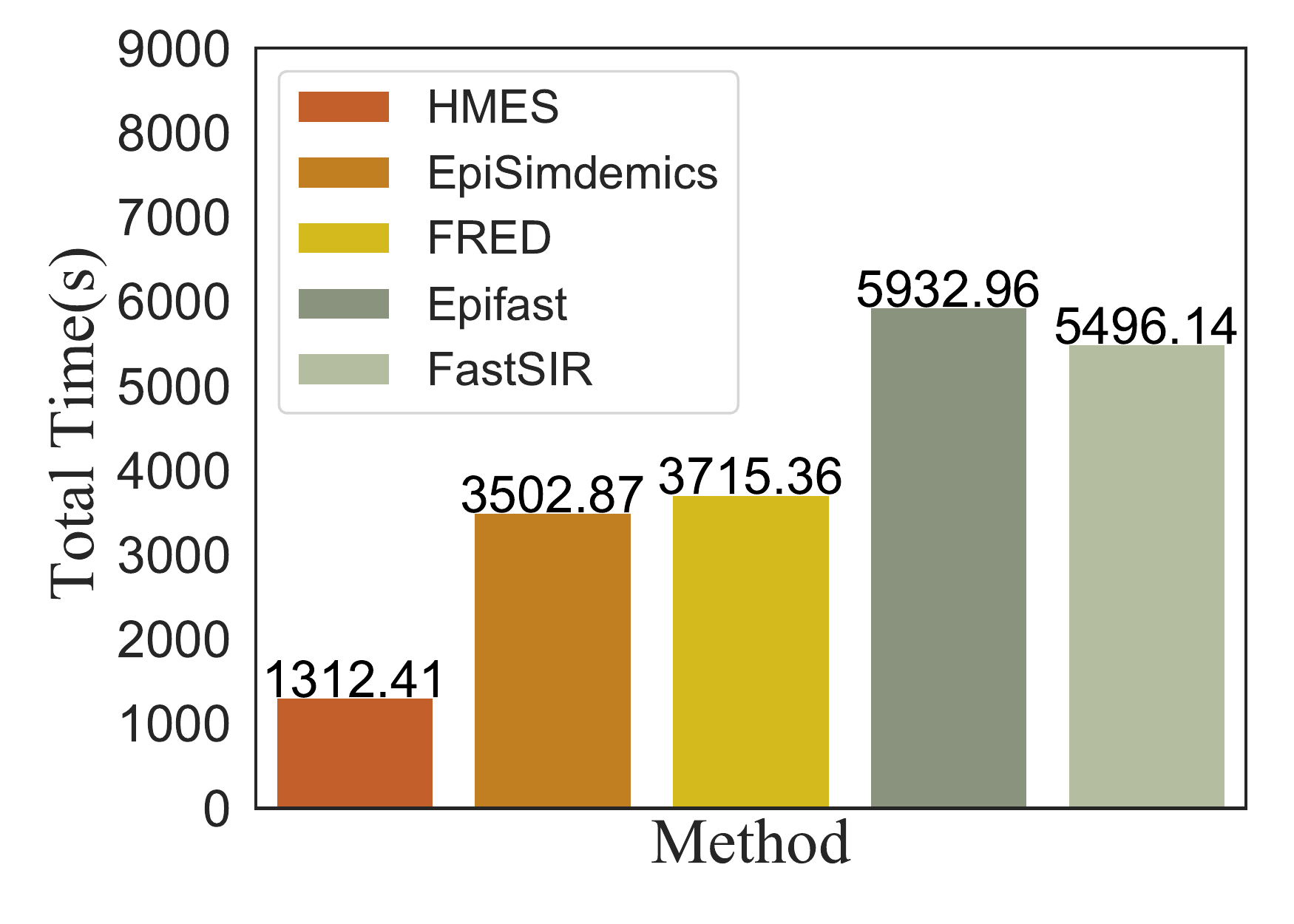} 
			\vspace{-0.2cm}
			\\
		\end{tabular}
		\caption{Total runtime w.r.t. population size, 10K(small county), 100K (small city), 1M (medium city), 10M (large city).  }
		\label{fig:comparison-population-scale}
	\end{figure*}
	
	\subsection{Proof of Proposition~5.1}
	We begin with the equivalence analysis of our Alg.~2(in main paper)
	and Alg.~\ref{alg:intervention-naive} as essentially they solve the vertex cover problem on dynamic bipartite graph. 
	
	{\bf Problem Formulation}
	At at time stamp $t'$, the snapshot is a bipartite graph consisting of nodes of people and nodes of locations. The adjacency list of location vertices is $C_{L,t'}$, i.e. the two sets of vertices are connected by $C_{L,t'}$. The task is to find a subset $C_{L}^{\star}$ such that it includes at least one endpoint of every edge of the bipartite graph. 
	
	{\bf Guarantee on Equivalence}
	In this sense, a risky person is traced or traversed if the person node is covered. As the set $C_L$ in  Alg.~2(of main paper)
	is essentially the collection of set $C_p$ over all potential people $\mathbf{B}$ in Alg.~\ref{alg:intervention-naive}, the solution of two algorithms are identical without tracing omissions.
	
	{\bf Guarantee on Minimum Necessary Set} 
	Given the potential risky locations $C_L$, as the solution of vertex cover in bipartite graph could be obtained trivially with solutions to maximum matching by traversing the uncovered location nodes in the augmenting path, while the maximum matching could be obtained by Hungarian algorithm \cite{kuhn1955hungarian} with $O(|\mathbf{C}_L| \cdot q)$ polynomial time without extra time complexity.

	\section{Additional Experiments}
	\begin{algorithm}[ht]  
		\caption{Basic Contact Tracing}  
		\label{alg:intervention-naive}  
		\KwIn{Infected people set $\mathbf{F^{(d)}}$, people's trajectories  $\mathbf{J}$, visiting people set of locations $\mathbf{V}$, current step $t$}
		\KwOut{Intervention strategy for next day $\mathbf{\lambda^{(d+1)}}$}
		\textbf{Parameters}: Tracking steps $\tau$,  maximum tracing order $maxOrder$, intervention type $\beta$\\
		Initialize potential infection source people set $\mathbf{B}
		\gets \mathbf{\mathbf{F^{(d)}}}$ \\
		\For{$ order = 1 \to maxOrder$}{
			Initialize concerned people set $\mathbf{C}_p \gets \{\}$ \\
			\For{$m \in \mathbf{B}$}{
				\For { $ t' = t-\tau \to t$}{
					Query the place that person $m$ visited at $t'$, $l = \mathbf{J}_m^{t'}$ \\
					\For {$m' \in \mathbf{V_l^{t'}} $}{
						$\mathbf{C}_p \gets \mathbf{C}_p + m'$
					}
				}
			}
			\For{$ c \in \mathbf{C}_p$}
			{ Impose intervention strategy $\mathbf{\lambda^{(d+1)}}[m] \gets \beta$} 
			Update $\mathbf{B}$ by $\mathbf{B} \gets \mathbf{C}_p$
		}
	\end{algorithm}  
	\subsection{Detailed Experimental Settings}
	\textbf{Setting for case study in Section~6.2}
	To test how well our simulation can fit the real data, we present our case study on COVID-19 pandemic with recent publicly available data from United States Census\textsuperscript{\ref{note:census}}
	We extracted necessary features from the census and synthesize virtual cities with various sizes, and adjust the infection rate in \ours to fit the real epidemic progression curve of cities with same scale. 
	
	\begin{table}[htbp]
		\centering
		\caption{Experiment Settings for Case Study}
		\label{tab:setting-case-study}
		\vspace{-0.1cm}
		\resizebox{0.47\textwidth}{!}{%
			\begin{tabular}{c|c|c|c|c|c}
				\toprule
				$p_{infection}$   & $p_{travel}$ & $p_{return}$ & intervention & $t_{cure}$ &$t_{isolate}$\\ \midrule
				0.0085       & 0.12    & 0.25    & hospitalize + isolate &7 &3\\
				\bottomrule
			\end{tabular}
		}
	\end{table}
	
	We set a fixed infection rate $p_{infection}$ for all the cities, and people have a certain probability $p_{travel}$ to travel to other cities , the specific city they travel to is decided by the distance (following a long-tail distribution). During travels, people have a certain probability $p_{return}$ to return back each day. Once a person is infected, he will be sent to hospital for $t_{cure}$ days, his contacts within two days will be isolated for $t_{isolate}$ days. Specific settings are shown in Table~\ref{tab:setting-case-study}.
	
	\textbf{Setting for case study in Section~6.3}
The infection data is publicly available\footnote{https://data.pa.gov/Health/COVID-19-Aggregate-Cases-Current-Daily-County-Heal}\footnote{https://dshs.texas.gov/coronavirus/AdditionalData.aspx}, and is taken as first thirty days after outbreak, where all the initial infected count is taken as 10. 
The population and tracts data are also public\textsuperscript{\ref{note:census}}. 


\subsection{Additional Experiment Results on Efficiency}\label{supp:experi}

{\bf Population size.} We demonstrate the impact of population size on the running time of different methods by varying the total population with population density unchanged, shown in Figure~\ref{fig:comparison-population-scale}. These three cases represent three typical city-scale in the world, a large city with 10M population and 10K locations, a medium city with 1M population and 1K locations, and a small town with 100K population and 100 locations. Under different populations, \ours outperforms all the baselines. The running time of \ours scales nearly linearly with the increasing population size, and performs fastest simulation under all settings. 


{\bf Population density.} We further compare the scalability in simulating different population densities.
The experiments are done on a city with 1 million population and various numbers of locations. As shown in Figure~\ref{fig:density-infection}(a), running time of \ours is the smallest under different settings and stays almost unchanged.  The other two location-based methods EpiSimdemics and FRED are also not sensitive to the population density, while the running time of EpiFast and FastSIR explode in high-density scenarios. \nop{Note that experiments for intervention on our single node computing platform even fails for high density on large scale for network-work based methods, and we estimate corresponding running time by taking intervention on proportion of people\zhengqing{is there a better saying?}. }
\begin{figure}[htbp]
	\centering 
	\subfigure[Total runtime w.r.t population density]{ 
		\includegraphics[width=0.22\textwidth]{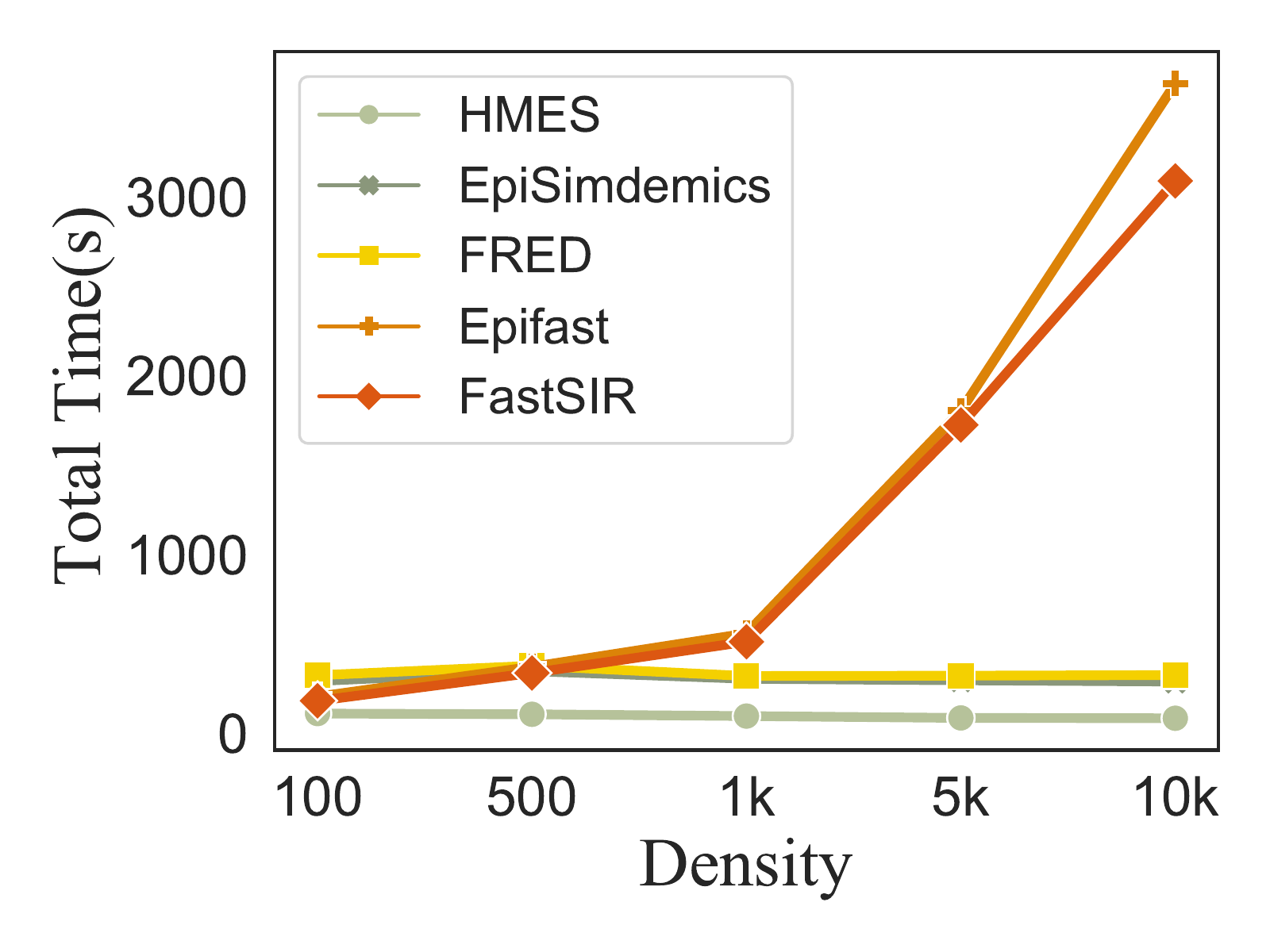}
	} 
	\subfigure[Total runtime w.r.t infection rate]{ 
		\includegraphics[width=0.22\textwidth]{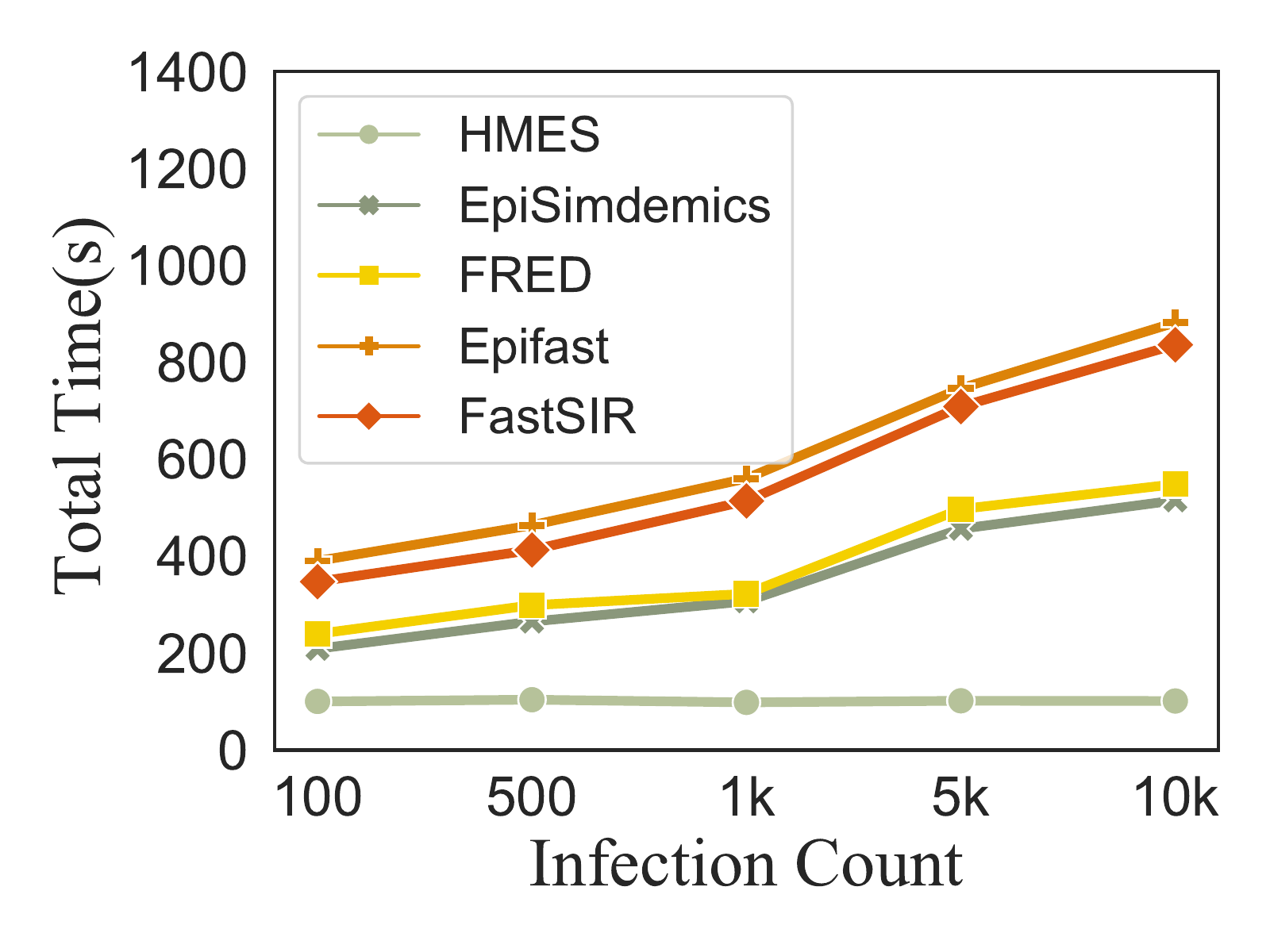}
	} 
	\vspace{-6mm}
	\caption{Scalability in population density \& infection rate}
	\label{fig:density-infection}
\end{figure}

{\bf Infection rate.}
We also test how these methods scale when simulating diseases with different infection rates. As shown in Figure~\ref{fig:density-infection}(b), \ours achieves the lowest running time under different infection rates and remains stable with various infection rate, while other methods become much slower as the infection rate increases. 

\end{document}